\documentclass[superscriptaddress,twocolumn,aps,pra,preprintnumbers,notitlepage]{revtex4-1}
\usepackage{amsmath,amssymb}
\usepackage{lipsum} 
\usepackage[latin9]{inputenc}
\usepackage{graphicx}
\usepackage{dcolumn}
\usepackage{bbold}
\usepackage{bm}
\usepackage[usenames]{xcolor}
\usepackage[colorlinks,bookmarks=false,citecolor=blue,linkcolor=blue,urlcolor=blue,hyperfootnotes=true]{hyperref}
\usepackage{mathtools}

\begin{document}

\title{Thermodynamic signatures of an antiferromagnetic quantum critical point inside a superconducting dome}

\author{Vanuildo S. de Carvalho}
\affiliation{School of Physics and Astronomy, University of Minnesota, Minneapolis, Minnesota 55455, USA}
\affiliation{Instituto de F\'{i}sica Gleb Wataghin, Unicamp, 13083-859, Campinas-SP, Brazil}
\author{Andrey V. Chubukov}
\affiliation{School of Physics and Astronomy, University of Minnesota, Minneapolis, Minnesota 55455, USA}
\author{Rafael M. Fernandes}
\affiliation{School of Physics and Astronomy, University of Minnesota, Minneapolis, Minnesota 55455, USA}
\date{\today}

\begin{abstract}
Recent experiments in unconventional superconductors, and in particular iron-based materials, have reported evidence of an antiferromagnetic quantum critical point (AFM-QCP) emerging inside the superconducting dome of the phase diagram. Fluctuations associated with such an AFM-QCP are expected to promote unusual temperature dependencies of thermodynamic quantities. Here, we compute the $T$ dependence of the specific heat $C(T)$ deep inside a fully gapped $s^{+-}$ superconducting state as the AFM-QCP is approached. We find that, at the AFM-QCP, the specific heat $C(T)$ vanishes quadratically with temperature, as opposed to the typical exponential suppression seen in fully-gapped BCS superconductors. This robust result is due to a non-analytic contribution to the free-energy arising from the general form of the bosonic (AFM) propagator in the SC state. Away from the AFM-QCP, as temperature is lowered, $C(T)$ shows a crossover from a $T^2$ behavior to an exponential behavior, with the crossover temperature scale set by the value of the superconducting gap and the distance to the QCP. We argue that these features in the specific heat can be used to unambiguously determine the existence of AFM-QCPs inside the superconducting domes of iron-based and other fully gapped unconventional superconductors.
\end{abstract}

\maketitle

\section{Introduction}\label{Sec_I}

Strongly correlated electronic systems often display complex phase diagrams in which the competition among different types of long-range order extends down to zero temperature \cite{Sachdev-PT(2011),Vojta-RPP(2003),Lohneysen-RMP(2007)}. In this situation, the system may display one or several quantum critical points (QCPs), which are continuous zero-temperature phase transitions that separate two distinct symmetry-broken ground states \cite{Sacdev-CUP(2011)}. In the phase diagram of many unconventional superconductors, the superconducting (SC) dome is often peaked near a putative antiferromagnetic (AFM) QCP \cite{Scalapino2012}. It has been widely
discussed that AFM quantum critical fluctuations can enhance $T_c$ and also lead to strange normal-state properties \cite{Abanov-AP(2003),Metlitski-PRBb(2010)}. However, experimentally identifying such an AFM-QCP is challenging. Ideally, one would suppress the SC dome, e.g., by applying a magnetic field, in order to reveal the underlying QCP \cite{Taillefer2019}. The very large values of the field that are necessary to kill $T_c$, and its impact on the magnetic state itself, make this a complicated task \cite{Analytis2016}.

In some materials, where the competition between AFM and SC is not too strong, however, the AFM transition line can persist even inside the SC dome \cite{Fernandes-PRB(2010),Vorontsov2010,Fernandes2013,Andersen2016,Barci2018,Senechal2019}, suggesting the presence of an AFM-QCP coexisting with long-range SC order (see Fig. \ref{FeSC}(a)). This is believed to be the case in some iron-based superconductors \cite{Julien2009,Bernhard2010,Johrendt2011,Ma2012,Yayu2013,Sky2018}, most prominently Ba(Fe$_{1-x}$Co$_x$)$_2$As$_2$ and BaFe$_2$(As$_{1-x}$P$_x$)$_2$ \cite{Hashimoto-S(2012),Matsuda-ARCMP(2014),Prozorov-NJP(2020)}, and in certain $f$-electron systems, such as CeCo(In$_{1-x}$Cd$_x$)$_5$ and Nd-doped CeRhIn$_5$ \cite{Nicklas2007,Rosa2017}. These systems offer the appealing possibility of probing an AFM-QCP without having to destroy the SC dome. Therefore, to unambiguously identify an AFM-QCP enclosed by a SC dome, it is fundamental to elucidate its manifestations on experimentally accessible quantities.

Recently, measurements of the zero-temperature penetration depth $\lambda(0)$ were employed to search for AFM quantum criticality inside the SC dome of the iron pnictides discussed above \cite{Hashimoto-S(2012),Matsuda-ARCMP(2014),Prozorov-NJP(2020)}. While a sharp peak in $\lambda(0)$ was observed as the SC dome was traversed, theoretically it remains unclear if such a feature can be uniquely attributed to an underlying AFM-QCP \cite{Chubukov-PRL(2013),Chowdhury-PRL(2013),Ikeda2013,Chowdhury2015,Dzero2015}. This motivates the study of how other observables, and in particular thermodynamic quantities, are affected by AFM fluctuations inside the SC dome.

In this paper, we determine the low-temperature behavior of the specific heat $C(T)$ of a fully gapped superconductor upon approaching an AFM-QCP. This QCP divides the SC dome in two regions: a pure SC state and a state where AFM and SC coexist microscopically (as opposed to phase-separate), as shown schematically in Fig. \ref{FeSC}(a). We specifically consider the case of $s^{+-}$ superconductivity, in which the gap changes sign between different bands \cite{Hirschfeld2011,Chubukov-ARCMP(2012)}. While such a state is ubiquitous in the iron pnictide materials, it has been proposed to be realized in other unconventional superconductors, such as CeCu$_2$Si$_2$ \cite{Yifeng2018}. Importantly, the fact that the spectrum is gapped allows us to perform controlled calculations and isolate the effects caused by quantum AFM fluctuations. Previously, it was shown that AFM fluctuations can significantly affect the specific heat jump at $T_c$ \cite{Varma2003,Vavilov-PRB(2014)}, which is in accordance with experimental measurements on BaFe$_2$(As$_{1-x}$P$_x$)$_2$ \cite{Carrington-PRL(2013)}. Our focus here is on the impact of AFM fluctuations on the low-temperature behavior of the specific heat, as the system approaches a QCP.

Our model consists of a multi-band, two-dimensional SC in proximity to an AFM-QCP. By computing the contribution of the AFM fluctuations to the SC free-energy, we find that, at the AFM-QCP, $C(T)$ vanishes as $T^2$ inside the fully gapped SC state. This is in sharp contrast to the behavior far away from the QCP, where $C(T)$ displays the standard exponential suppression $e^{-|\Delta|/T}$, where $|\Delta|$ is the zero-temperature SC gap. This power-law (as opposed to exponential) behavior of the specific heat comes from a non-analytic contribution of the soft AFM fluctuations to the free energy of the SC state. It is a robust result rooted on the general form of the AFM propagator inside the SC state, which resembles the propagator of the AFM fluctuations of a two-dimensional quantum Heisenberg model obtained by Chubukov, Sachdev, and Ye (CSY) within the large-$N$ approach \cite{Chubukov-PRB(1994)}. Away from the AFM-QCP, the specific heat displays a crossover behavior. In particular, the $T^2$ behavior crosses over to the more typical exponential behavior at a temperature $T^{*}$ of the order of $\sqrt{r} |\Delta|$, where $r$ measures the distance to the AFM-QCP located at $r=0$. Our results provide a simple diagnostics to identify AFM-QCPs inside the SC domes of unconventional superconductors.

The paper is structured as follows. In Sec. \ref{Sec_II}, we briefly describe the multi-band model employed here and comment on its applicability to address the physical properties of iron-based compounds. Section \ref{Sec_III} deals with the evaluation of the AFM propagator inside the superconducting state. In Sec. \ref{Sec_IV}, we derive the free energy in the vicinity of the AFM quantum phase transition and compute the specific heat. In Sec. \ref{Sec_V}, we analyze  how the specific heat is affected by  the fermion-induced mode-mode coupling between AFM fluctuations and discuss the relation of our results to earlier study of the specific heat in the non-linear $\sigma$-model near a critical coupling (Ref. \cite{Chubukov-PRB(1994)}). Finally, Sec. \ref{Sec_VI} contains our conclusions and discussions of our findings in connection with experiments.

\section{Microscopic Model}\label{Sec_II}

\begin{figure}[t]
\centering
\includegraphics[width=1.0\linewidth]{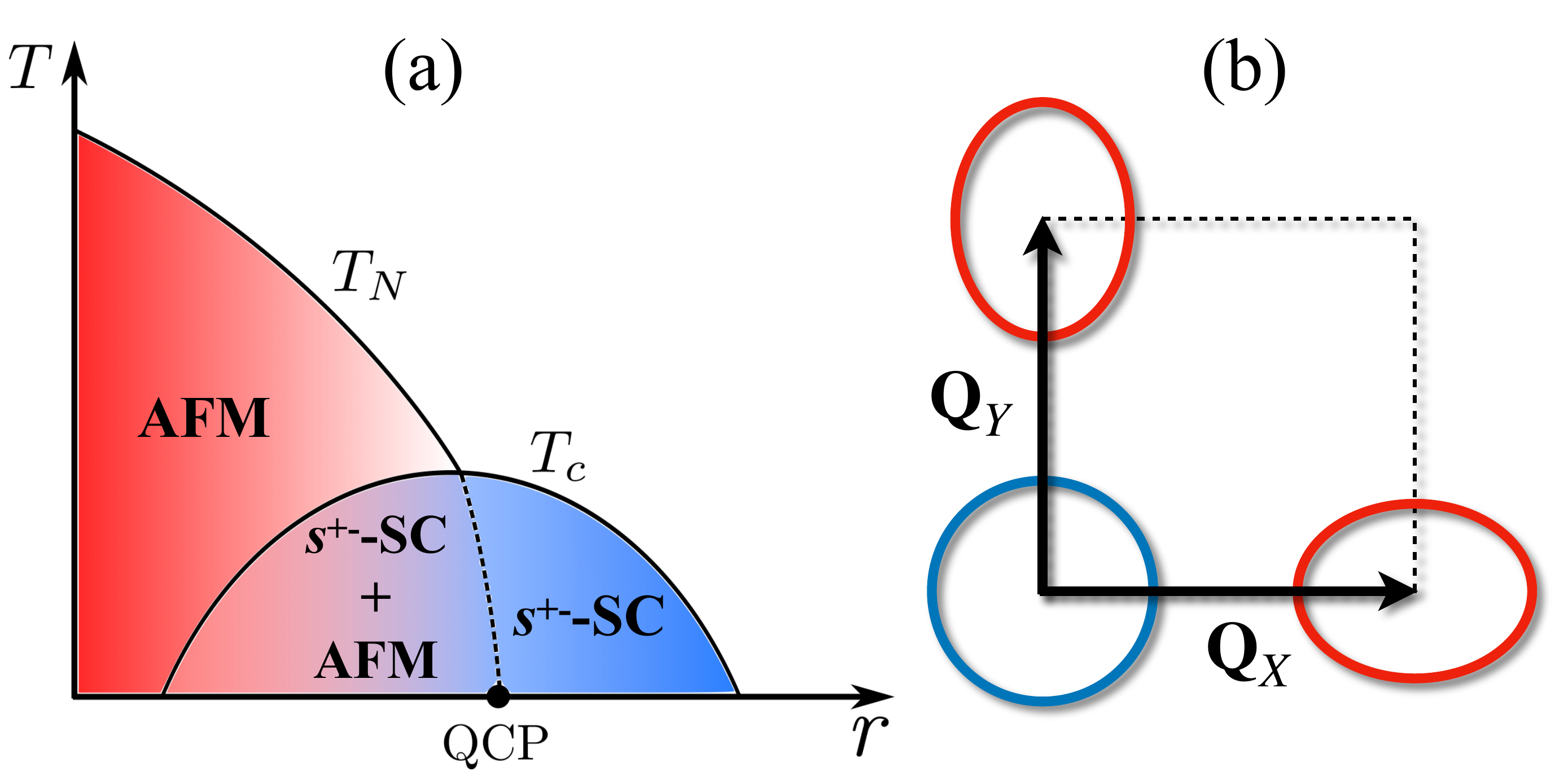}
\caption{(a) Schematic phase diagram showing AFM and $s^{+-}$-SC orders with critical temperatures $T_N$ and $T_c$, respectively. As some control parameter $r$, representing for example doping, is increased, the AFM temperature $T_N$ decreases and eventually becomes zero at a putative QCP at $r=0$. As occurs in some iron-based compounds, the QCP is located inside the SC dome, such that both AFM and $s^{+-}$-SC orders exhibit microscopic coexistence for a finite range of $r$. Note that the $T_N$ line may display a back-bending inside the SC dome (not shown here), as seen in certain iron-based materials \cite{Fernandes-PRB(2010)}. (b) Simplified band structure depicting a circular hole-like Fermi pocket and two elliptical electron-like Fermi pockets displaced from the hole pocket by the AFM wave-vectors $\mathbf{Q}_X = (\pi,0)$ and $\mathbf{Q}_Y = (0,\pi)$.}\label{FeSC}
\end{figure}

Our main conclusions are a direct consequence of the analytical form of the AFM propagator inside a nodeless superconducting state, and are thus independent of microscopic considerations. Yet, it is instructive to derive such a propagator from a known microscopic model. Since iron-based superconductors (FeSCs) are the main candidate to display an AFM-QCP inside a fully gapped SC state, we consider a simple three-band model, which has been widely investigated previously \cite{Chubukov-ARCMP(2012),Fernandes-RPP(2017)}. In momentum space, the non-interacting part of the three-band Hamiltonian is given by
\begin{align}
\mathcal{H}_0=&\sum_{\mathbf{k},\sigma}(\varepsilon_{c,\mathbf{k}}-\mu)c^\dagger_{\mathbf{k},\sigma}c_{\mathbf{k},\sigma}\nonumber\\
&\hspace{1.0cm}+\sum_{\mathbf{k},\sigma, a}(\varepsilon_{d,a,\mathbf{k}+\mathbf{Q}_a}-\mu)d^\dagger_{a, \mathbf{k}+\mathbf{Q}_a,\sigma}d_{a, \mathbf{k}+\mathbf{Q}_a,\sigma},
\end{align}
where $c^\dagger_{\mathbf{k},\sigma}$  and $d^\dagger_{a, \mathbf{k},\sigma}$ are, respectively, the creation operators for hole-like and electron-like excitations with spin projection $\sigma\in\{\uparrow,\downarrow\}$ and momentum $\mathbf{k}$. The index $a=X,Y$ labels the two symmetry-related electron pockets. In particular, while the hole pocket is centered at the $\Gamma=(0,0)$ point of the Brillouin zone, the electron pockets are centered at $X= (\pi,0)$ and $Y= (0,\pi)$, corresponding to $a = X$ and $a= Y$, respectively. The dispersions of the hole and electron bands are parametrized according to $\varepsilon_{c,\mathbf{k}}=\varepsilon_{c,0}-\mathbf{k}^2/(2m)$ and $\varepsilon_{d,a,\mathbf{k}+\mathbf{Q}}=-\varepsilon_{d,0}+k^2_x/(2m_a)+k^2_y/(2m_{\bar{a}})$, where we introduced the notation $\bar{a} = Y,X$ for $a = X,Y$. Here, $\varepsilon_{c,0}$ and $\varepsilon_{d,0}$ are energy offsets, and $m$, $m_x$, and $m_y$ are band masses. As schematically depicted in Fig. \ref{FeSC}(b), $\varepsilon_{c,\mathbf{k}}$ describes a circular hole pocket at the center of the Brillouin zone, whereas $\varepsilon_{d,a, \mathbf{k}+\mathbf{Q}_a}$ gives elliptical electron pockets displaced from the hole pocket by the AFM wave-vectors $\mathbf{Q}_X = (\pi,0)$ and $\mathbf{Q}_Y = (0,\pi)$.

The interacting part of the three-band model Hamiltonian in band basis contains four-fermion couplings that can be classified as density-density inter- and intra-pocket interactions, exchange inter-pocket interaction, and pair-hopping inter-pocket interaction \cite{Chubukov-PRB(2008)}. A renormalization-group (RG) analysis of this model reveals two main instabilities in the phase diagram \cite{Chubukov-ARCMP(2012),Fernandes-RPP(2017)}: an AFM phase with ordering vector $\mathbf{Q}_X = (\pi,0)$ or $\mathbf{Q}_Y = (0,\pi)$ and an $s^{+-}$ SC state, in which the hole pocket has a uniform gap with opposite sign as the uniform gaps in the electron pockets. While at low doping levels (tuned by the chemical potential in the model) AFM wins, at intermediate doping levels the $s^{+-}$ SC state wins. This gives rise to the possibility of an AFM-QCP inside the SC dome. Importantly, it has been argued that only in the case of an $s^{+-}$ SC state a microscopic coexistence between AFM and SC is possible \cite{Fernandes-PRB(2010)}.

As discussed in Ref. \cite{Fernandes-RPP(2017)}, to capture the main properties of the interplay between AFM and SC, it is convenient to further simplify the model and focus on the interaction between the hole pocket and only one of the two electron pockets. Hereafter, we will thus consider this simplified two-band model, dropping the index $a$ and referring to the AFM wave-vector simply as $\mathbf{Q}$. To proceed, based on the RG results, we project the interacting Hamiltonian of the two-band model into the two leading instabilities, AFM and SC:
\begin{equation}\label{Eq_afm_Ham}
\mathcal{H}_\text{AFM}=-\frac{g_{\mathrm{afm}}}{2}\sum_{j}\big(c^\dagger_{j,\alpha}\boldsymbol{\sigma}_{\alpha,\alpha'}d_{j,\alpha'}\big)\cdot\big(d^\dagger_{j,\beta}\boldsymbol{\sigma}_{\beta,\beta'}c_{j,\beta'}\big),
\end{equation}
where $g_{\mathrm{afm}}>0$ is the coupling constant in the AFM channel and:
\begin{equation}\label{Eq_SC_Ham}
\mathcal{H}_\text{SC}=\frac{g_{\mathrm{sc}}}{2}\sum_{j}\big[ c^\dagger_{j,\alpha}(i\sigma^y)^\dagger_{\alpha,\alpha'}c^\dagger_{j,\alpha'}d_{j,\beta}(i\sigma^y)_{\beta,\beta'}d_{j,\beta'}+\text{H.c.}\big],
\end{equation}
with $g_{\mathrm{sc}}>0$ being the coupling constant in $s^{+-}$ SC channel.

\section{AFM propagator in the SC state}\label{Sec_III}

We first perform the Hubbard-Stratonovich decoupling of the interacting Hamiltonians $\mathcal{H}_\text{AFM}$ and $\mathcal{H}_\text{SC}$ by introducing the order parameters $\mathbf{M}_{j}=g_{\mathrm{afm}}\langle c^\dagger_{j,\sigma}\boldsymbol{\sigma}_{\sigma,\sigma'}d_{j,\sigma'}\big\rangle$, $\Delta_{c,j}=g_{\mathrm{sc}}\langle c^\dagger_{j,\sigma}(i\sigma^y)_{\sigma,\sigma'}c^\dagger_{j,\sigma'}\rangle$, and $\Delta_{d,j}=g_{\mathrm{sc}}\langle d^\dagger_{j,\sigma}(i\sigma^y)_{\sigma,\sigma'}d^\dagger_{j,\sigma'}\rangle$ for AFM and SC orders, respectively. We consider that the system is on the magnetically disordered side of the QCP and restrict the temperature scale to the regime $T\ll T_c$. In this situation, the SC gaps $\Delta_c(T)$ and $\Delta_d(T)$ do not vary with doping and are, approximately, equal to their ground-state values \cite{Fernandes-PRB(2010)}, i.e., $\Delta_c(T)=-\Delta_d(T)=\Delta$. Since in this situation the fermionic excitations become gapped, they can be formally integrated out to obtain the effective AFM action. By considering the Gaussian fluctuations of the AFM order parameter around the mean-field solution, we find that the contribution to the action from AFM fluctuations inside the SC state gives
\begin{equation}\label{afm_fluct_FreeEnergy}
\delta\mathcal{S}[\Delta]=\frac{1}{2}T\sum_{\mathbf{q},\Omega_m}\delta\bar{M}^a_{\mathbf{q},i\Omega_m}\bigg[4\frac{\delta_{ab}}{g_{\mathrm{afm}}}-\Pi_{ab}(\mathbf{q},i\Omega_m)\bigg]\delta M^b_{\mathbf{q},i\Omega_m},
\end{equation}
where $\delta M^b_{\mathbf{q},i\Omega_m}$ denotes AFM (quantum) fluctuations, $\mathbf{q}$ is the momentum deviation from the AFM wave-vector $\mathbf{Q}$ (either ${\bf Q}_X$ or ${\bf Q}_Y$), $\Omega_m=2\pi m T$ is a bosonic Matsubara frequency, and $\Pi_{ab}(\mathbf{q},i\Omega_m)$ refers to the particle-hole bubble defined according to
\begin{align}
&\hspace{-0.3cm}\Pi_{ab}(\mathbf{q},i\Omega_m)\nonumber\\
&\hspace{-0.35cm}=-T\sum_{\omega_n}\int_\mathbf{k}\operatorname{tr}[\varSigma^a\mathcal{G}(\mathbf{k},i\omega_n)\varSigma^b\mathcal{G}(\mathbf{k}+\mathbf{q},i\omega_n+i\Omega_m)],
\end{align}
Here, $\int_\mathbf{k}(\cdots)\equiv \int_\text{BZ}\frac{d^2\mathbf{k}}{(2\pi)^2}(\cdots)$ is the integral over the Brillouin zone, and the matrix Green's function is given by:
\begin{align}
\hspace{-0.3cm}\mathcal{G}(\mathbf{k},i\omega_n)&=\bigg[\begin{pmatrix} 1 & 0 \\ 0 & 0 \end{pmatrix}\otimes\begin{pmatrix} i\omega_n-\xi_{c,\mathbf{k}} & -i\sigma^y\Delta^* \\ i\sigma^y\Delta & i\omega_n+\xi_{c,-\mathbf{k}} \end{pmatrix}\nonumber\\
&+\begin{pmatrix} 0 & 0 \\ 0 & 1 \end{pmatrix}\otimes\begin{pmatrix} i\omega_n-\xi_{d,\mathbf{k}} & i\sigma^y\Delta^* \\ -i\sigma^y\Delta & i\omega_n+\xi_{d,-\mathbf{k}} \end{pmatrix}\bigg]^{-1},
\end{align}
where $\xi_{c,\mathbf{k}}=\varepsilon_{c,\mathbf{k}}-\mu$, $\xi_{d,\mathbf{k}}=\varepsilon_{d,\mathbf{k}+\mathbf{Q}}-\mu$, and $\boldsymbol{\varSigma}=\sigma^x\otimes\begin{pmatrix} \boldsymbol{\sigma} & \mathbb{0} \\ \mathbb{0} & -\boldsymbol{\sigma}^T \end{pmatrix}$ is an $8\times8$ matrix.

In order to evaluate the AFM propagator $\chi^{-1}_{ab}(\mathbf{q},i\Omega_m)=4\delta_{ab}/g_{\mathrm{afm}}-\Pi_{ab}(\mathbf{q},i\Omega_m)$ that enters in the action $\delta\mathcal{S}[\Delta]$, we follow Refs. \cite{Vavilov-SST(2010),Vorontsov2010} and first write the dispersions $\xi_{c,\mathbf{k}}$ and $\xi_{d,\mathbf{k}}$ for the hole- and electron-like bands as
$\xi_{d,\mathbf{k}}=-\xi_{c,\mathbf{k}}+\delta_\varphi$, $\xi_{c,\mathbf{k}+\mathbf{q}}=\xi_{c,\mathbf{k}}+\delta_\mathbf{q}$,
where $\delta_\varphi=\delta_0+\delta_2\cos(2\varphi)$ and $\delta_\mathbf{q}=v_F q\cos(\varphi-\theta)$. Here, $\varphi$ and $\theta$ are the angles that the vectors $\mathbf{k}$ and $\mathbf{q}$ make with the $\hat{\mathbf{x}}$ axis, respectively. The parameters $\delta_0$ and $\delta_2$ are given by $\delta_0=\varepsilon_{c,0}-\varepsilon_{d,0}-2\mu+k^2_F(m^{-1}_x+m^{-1}_y-m^{-1}/2)/4$ and $\delta_2=k^2_F(m^{-1}_x-m^{-1}_y)/4$; they measure, respectively, the offset energy between the hole and electron pockets and the ellipticity of the electron band. Previous calculations have shown that this model admits a transition from a SC state to a SC-AFM coexistence state over a certain parameter range \cite{Vavilov-SST(2010),Vorontsov2010}.

We assume as usual that the most relevant contribution to $\Pi_{ab}(\mathbf{q},i\Omega_m)$ comes from the electronic states close to the Fermi surface, such that $\int_\text{BZ}\frac{d^2\mathbf{k}}{(2\pi)^2}(\cdots)=\nu_0\int^{2\pi}_0\frac{d\varphi}{2\pi}\int^{\infty}_{-\infty}d\xi(\cdots)$, where $\nu_0$ is the density of states at the Fermi level. Moreover, due to spin-rotational symmetry, $\Pi_{ab}(\mathbf{q},i\Omega_m)=\delta_{ab}\Pi(\mathbf{q},i\Omega_m)$. Expanding $\Pi(\mathbf{q},i\Omega_m)$ for small momentum and frequency, we have:
\begin{equation}
\Pi(\mathbf{q},i\Omega_m)=\Pi(0,0)+\delta\Pi(\mathbf{q},0)+\delta\Pi(0,i\Omega_m),
\end{equation}
where the last two terms are, respectively, the leading-order momentum and frequency contributions. The first term combines with the constant term in Eq. \eqref{afm_fluct_FreeEnergy} to give $r\equiv\nu^{-1}_0[4/g_{\mathrm{afm}}-\Pi(0,0)]$, which measures the distance to the AFM transition. Evaluating the particle-hole bubble gives
\begin{equation}\label{Eq_AFM_mass}
r=4\bigg[\frac{1}{\nu_0g_{\mathrm{afm}}}-2\pi T\sum_{\omega_n>0}\bigg\langle\frac{E_n}{E^2_n+(\delta_\varphi/2)^2}\bigg\rangle_\varphi\bigg],
\end{equation}
where $E_n \equiv \sqrt{\omega^2_n+|\Delta|^2}$ is the spectrum in the SC phase and $\langle(\cdots)\rangle_\varphi \equiv \int^{2\pi}_{0}\frac{d\varphi}{2\pi}(\cdots)$ denotes the angular average around the Fermi surface. At $T=0$ one obtains:
\begin{equation}
r(\delta,\Delta)=4\log\bigg(\frac{T_{c,0}}{T_{N,0}}\bigg)+\bigg\langle\frac{2|\delta_\varphi|\cosh^{-1}\Big(\sqrt{1+\frac{\delta^2_\varphi}{4|\Delta|^2}}\Big)}{\sqrt{|\Delta|^2+(\delta_\varphi/2)^2}}\bigg\rangle_\varphi,
\end{equation}
where we made the dependence of $r$ on the set of parameters $\delta=\{\delta_0,\delta_2\}$ and the SC gap $\Delta$ explicit, as they determine the position of the AFM-QCP. In this expression, $T_{N,0}$ refers to the transition temperature to the pure AFM state at perfect nesting (i.e. $\delta=0$) and $T_{c,0}$ is the transition temperature to the SC state in the absence of AFM order.

Evaluating the Matsubara sums in both $\delta\Pi(\mathbf{q},0)$ and $\delta\Pi(0,i\Omega_m)$ at $T=0$, we obtain
\begin{align}
&\delta\Pi(\mathbf{q},0)=-\nu_0\eta(\delta,\Delta,\theta)(v_F q)^2,\\
&\delta\Pi(0,i\Omega_m)=-\nu_0\kappa(\delta,\Delta)\Omega^2_m,
\end{align}
where
\begin{align}
&\eta(\delta,\Delta,\theta)=\frac{1}{2}\bigg\langle\cos^2(\varphi-\theta)\bigg\lbrace\frac{2|\Delta|^2-(\delta_\varphi/2)^2}{[|\Delta|^2+(\delta_\varphi/2)^2]^2}\nonumber\\
&\hspace{1.7cm}-\frac{3|\Delta|^2|\delta_\varphi|\cosh^{-1}\Big(1+\frac{\delta^2_\varphi}{2|\Delta|^2}\Big)}{4[|\Delta|^2+(\delta_\varphi/2)^2]^{5/2}}\bigg\rbrace\bigg\rangle_\varphi,\\
&\kappa(\delta,\Delta)=\frac{1}{2}\bigg\langle\frac{1}{|\Delta|^2+(\delta_\varphi/2)^2}+\frac{|\Delta|^2}{|\delta_\varphi|[|\Delta|^2+(\delta_\varphi/2)^2]^{3/2}}\nonumber\\
&\hspace{1.4cm}\times\cosh^{-1}\bigg(1+\frac{\delta^2_\varphi}{2|\Delta|^2}\bigg)\bigg\rangle_\varphi. \label{aux}
\end{align}
As a result, we find that the AFM propagator is
 $\chi_{ab}(\mathbf{q},i\Omega_m)=\delta_{ab}\chi(\mathbf{q},i\Omega_m)$, with $\chi(\mathbf{q},i\Omega_m)$ given by:
\begin{equation}\label{propagator}
\chi(\mathbf{q},i\Omega_m)=\frac{\nu^{-1}_0}{r(\delta,\Delta)+\eta(\delta,\Delta,\theta)(v_F q)^2+\kappa(\delta,\Delta)\Omega^2_m}.
\end{equation}
The dependence of $\eta(\delta,\Delta,\theta)$ on $\theta$ implies that the prefactors for $q^2_x$ and $q^2_y$ are actually different. This is because   ${\bf q}$ is a deviation from either ${\bf Q}_X$ or ${\bf Q}_Y$.

Two comments are in other at this point. First, a complementary approach  would be to treat the static $\chi (0,0)$ as an input,  introduce the coupling $g_{\text{afm}}$ between collective spin fluctuations and low-energy fermions in a superconductor, and treat the polarization $\Pi (\mathbf{q}, i\Omega_m)$ as a bosonic self-energy.  The full $\chi (\mathbf{q}, i\Omega_m)$  in this approach has the same form as in Eq. \eqref{propagator}, but $\eta$ and $\kappa$ acquire additional factors $(\nu_0 g_{\text{afm}})^2$.  The results for the specific heat (see below) are identical in the two approaches, except for different powers of $\nu_0 g_{\text{afm}}$ in the crossover scales.

Second, the inverse of $\chi ({\bf q}, i \Omega_m)$ from Eq. \eqref{propagator} is the prefactor for the $(\delta M)^2$ term in the Ginzburg-Landau functional (see Eq. \eqref{afm_fluct_FreeEnergy}).  There exist Gaussian corrections to $r(\delta, \Delta)$ from the $b (\delta M)^4$ term in the functional (the mode-mode coupling term).  We proceed in Sec. \ref{Sec_IV} without including these corrections, and in Sec. \ref{Sec_V} we analyze how they affect the results.  The reasoning to first neglect the $b (\delta M)^4$ term is that in our case $b \sim |\Delta|/E_F$ is small, and mode-mode coupling affects the specific heat only at the smallest $T$  and smallest deviations from an AFM-QCP.

 \section{Thermodynamic properties in the vicinity of the AFM-QCP}\label{Sec_IV}

The AFM propagator in Eq. \eqref{propagator} is not particular to the microscopic model derived here, but is expected to describe a generic AFM-QCP inside a fully gapped SC state. This is because the gap in the spectrum eliminates the Landau damping typically present in a metallic AFM due to the decay of the AFM fluctuations in particle-hole excitations \cite{Millis-PRB(1993),Abanov-AP(2003),Chubukov-PRL(2004),Metlitski-PRBb(2010)}, resulting in the dynamic exponent $z=1$. Thus, for the remainder of the paper, to emphasize the generality of our results, we will omit the explicit dependence of the parameters $r$, $\eta$, and $\kappa$ on $\delta$ and $\Delta$. We will make use of the general result that $\kappa \sim |\Delta|^{-2}$, as can be verified in our model from Eq. \eqref{aux} (provided that $\delta_0,\delta_2$ are not much larger than $|\Delta|$). As for the dependence  of $\eta$ on the angle $\theta$, it is convenient to introduce $\eta_{x,y}$ such that $\eta(\theta)\equiv\eta_x \cos^2\theta+\eta_y \sin^2\theta$.

To obtain the AFM-fluctuations contribution to the free energy, we integrate out the AFM fluctuations in the action $\delta\mathcal{S}[\Delta]$ [see Eq. \eqref{afm_fluct_FreeEnergy}]. As a result, we obtain the free energy
from fluctuations near ${\bf Q}_X$ or ${\bf Q}_Y$ in the form:
\begin{equation}\label{F_final}
F(T)=NT\sum_{\Omega_m}\int_\mathbf{q}\log\left[\chi^{-1}(\mathbf{q},i\Omega_m)\right],
\end{equation}
where $N=3$ is the number of components of the AFM order parameter.

\begin{figure*}[t]
\centering \includegraphics[width=0.47\linewidth]{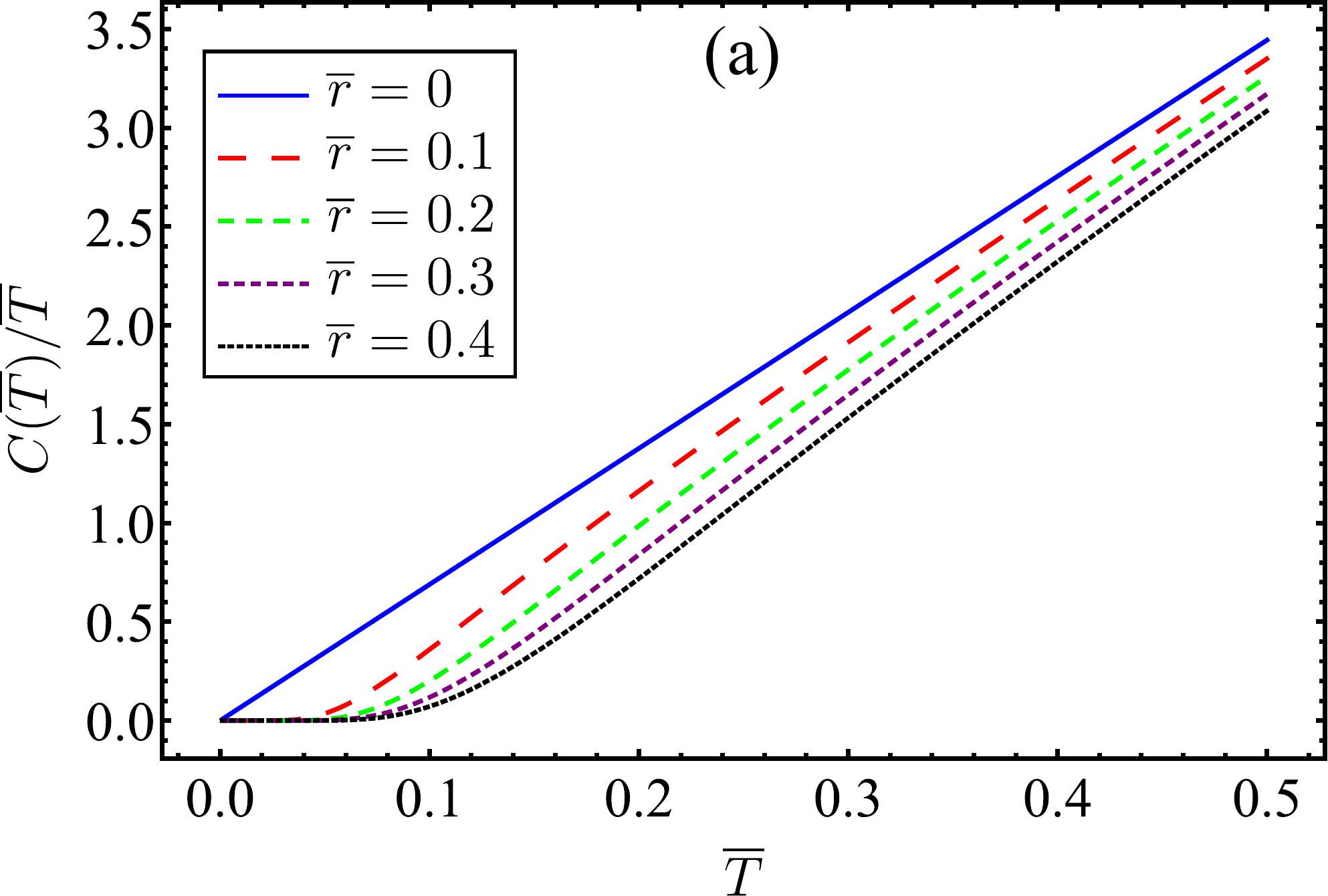}\hfill{}\includegraphics[width=0.46\linewidth]{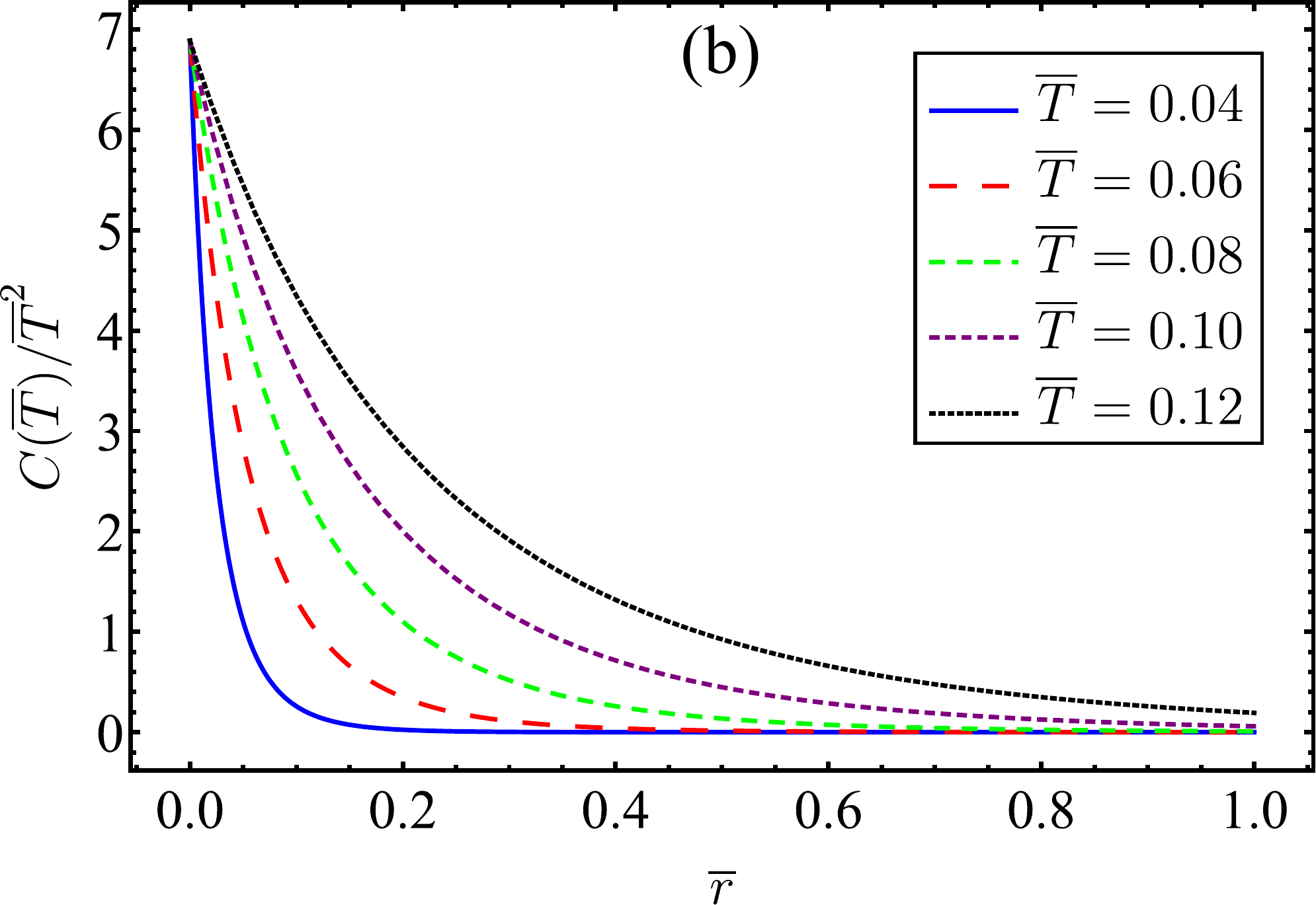}
\caption{(a) Dependence of the specific heat $C(\overline{T})/\overline{T}$ [in units of $\kappa |\Delta|^2/(\sqrt{\eta_x \eta_y}v^2_F)$] on the reduced temperature $\overline{T} \equiv T/|\Delta|$ and on the distance $\overline{r} \equiv r/\overline{\kappa}$ from the QCP measured in terms of the dimensionless parameter $\bar{\kappa}\equiv|\Delta|^2\kappa$. For temperatures of the order of $T^{*} \sim \sqrt{r} |\Delta|$, $C(T)\propto T^2$ even when the system is away from the quantum critical point. As the temperature is lowered, the $T^2$ behavior only persists down to $T=0$ at the QCP. For finite $r$, the specific heat is suppressed exponentially to zero for $T\ll T^{*}$. As shown in panel (b), $C(\overline{T})/\overline{T}^2$ increases monotonically upon approaching the QCP.}\label{Specific_Heat}
\end{figure*}

Like we said, in this section we neglect corrections to $r (\delta, \Delta)$ from mode-mode coupling and treat $r (\delta, \Delta) =r$ as a temperature-independent input parameter that measures the distance to a QCP. Evaluating the Matsubara sum on the right-hand side of Eq. \eqref{F_final} by the $\zeta$-function regularization method and integrating it over the two-dimensional momentum space, we obtain
\begin{equation}\label{Eq_Free_Energy}
F(T)=F_0-\frac{N\kappa T^3}{\pi\sqrt{\eta_x \eta_y}v^2_F}\Upsilon\bigg[\frac{r}{\kappa T^2}\bigg],
\end{equation}
where $F_0 \equiv N\int\frac{d^2\mathbf{q}}{(2\pi)^2}E_\mathbf{q}$ is a temperature-independent term defined as the momentum integral of the energy dispersion $E_\mathbf{q}=\sqrt{\frac{r}{\kappa}+\frac{\eta(\theta)}{\kappa}(v_Fq)^2}$, $\Upsilon(z)\equiv\sqrt{z}\operatorname{Li}_2(e^{-\sqrt{z}})+\operatorname{Li}_3(e^{-\sqrt{z}})$, and $\operatorname{Li}_s(z)$ are polylogarithms of order $s$.

At a finite distance from the QCP and small $T$, $r > \kappa T^2$, the temperature dependence of the magnetic part of the free energy is exponential:
$F_0 -F(T)  \sim T^2 \sqrt{r} e^{-\sqrt{r/\kappa T^2}}$, and we remind that $\kappa \sim 1/\Delta^2$ when
$\delta_0, \delta_2 \leq \Delta$.
At the magnetic QCP, $r=0$. Here we obtain
\begin{equation}
F_{\mathrm{QCP}}(T)  = F_0-\frac{N\zeta(3)\kappa}{\pi\sqrt{\eta_x \eta_y}v^2_F}T^3,
\end{equation}
where $\zeta(3)\approx 1.20205$ is the Ap\'{e}ry's constant. Thus, at the QCP, the free energy acquires a power-law dependence on the temperature, despite the fact that the ground state is a fully-gapped superconducting state. Note that the details of the microscopic model only enter this expression in the pre-factor via the coefficients $\eta_x$, $\eta_y$, $v_F$, and $\kappa$; in contrast, the $T^3$ dependence is universal. In this regard, for our original three-band model, the final expression would have to include contributions from fluctuations around both ${\bf Q}_X$ and ${\bf Q}_Y$ simultaneously. The determination of the pre-factor in this case is a bit more involved as compared to the two-band case, because the ratio between the gaps in the hole and in the electron pockets is not exactly $-1$. However, as explained, this does not affect our main result on the temperature dependence of $F$.

In fact, the emergence of the universal power-law dependence at $r=0$ can be traced to the non-analytic form of the frequency summand in $F(T)$. Indeed, after the momentum integration, we have at $r=0$,
\begin{equation}\label{F_aux}
F(T) - F_0 \propto T\sum_{m} \Omega_m^2 \log\left( \frac{\Lambda}{|\Omega_m|} \right)
\end{equation}
where $\Lambda \sim (\nu_0 \kappa)^{-1}$. Because of the $\log{|\Omega_m|}$ term, the frequency sum contains a universal $T^3$ contribution, which comes from $\Omega_m = 2\pi m T = O(T)$, i.e., from Matsubara numbers $n = O(1)$ (see Refs. \cite{Chubukov-PRB(1994),Chubukov-PRB(2005)} for details).

The same result for the magnetic contribution to the free energy, Eq. \eqref{F_final},
is obtained if we use the full expression for the free energy of a superconductor near a magnetic instability \cite{LW_1960,Eliashberg_1960,Haslinger2003}.
The bosonic part of the latter contains an additional $\sum_{q} \Pi \chi$ term,  but this term cancels out
by the contribution from closed linked skeleton diagrams \cite{Chubukov-PRB(2005)}.

Using Eq. \eqref{Eq_Free_Energy}, it is straightforward to calculate the specific heat $C(T) = -T \partial^2F(T)/\partial T^2$:
\begin{equation}\label{Eq_SHeat_01}
C(T)=\frac{6N\kappa T^2}{\pi\sqrt{\eta_x \eta_y}v^2_F}\mathcal{K}\bigg[\frac{r}{\kappa T^2}\bigg],
\end{equation}
where the function $\mathcal{K}(z)$
is expressed in terms of the polylogarithm function as:
\begin{align}
\mathcal{K}(z)\equiv&\;\sqrt{z}\operatorname{Li}_2(e^{-\sqrt{z}})+\operatorname{Li}_3(e^{-\sqrt{z}})+\frac{z^{3/2}}{6(e^{\sqrt{z}}-1)}\nonumber\\
&+\frac{z}{2}[\sqrt{z}-\log(e^{\sqrt{z}}-1)].
\end{align}
At the QCP, $r=0$, it follows from Eq. \eqref{Eq_SHeat_01} that the specific heat becomes quadratic in temperature:
\begin{equation}\label{Eq_SHeat_02}
C_{\mathrm{QCP}}(T) = \frac{6N\zeta(3)\kappa}{\pi\sqrt{\eta_x \eta_y }v^2_F}T^2.
\end{equation}
This behavior differs fundamentally from the one obtained either in a conventional SC \cite{Bardeen-PR(1957)} or in a two-dimensional metal close to an AFM-QCP without the presence of SC order \cite{Millis-PRB(1993),Abanov-AP(2003),Chubukov-PRL(2004),Metlitski-PRBb(2010)}. In fact, in the former case the specific heat is exponentially suppressed, $C(T)\sim T^{-3/2}e^{-|\Delta|/T}$, whereas in the latter case it becomes linear-in-$T$ with $C(T)\sim T \log(1/r)$. We note by passing that the $T \Omega^2_m \log{|\Omega_m|}$ term in the free energy also emerges in an interacting Fermi liquid in two dimensions and gives rise to $T^2$ non-analytic correction to the specific heat \cite{Chubukov-PRB(2005)}.

 As the system moves away from the AFM-QCP towards the magnetically disordered state (i.e. the overdoped side of the phase diagram), a new temperature scale $T^{*} \equiv \sqrt{r/ \kappa} \sim \sqrt{r}|\Delta|$ becomes important. For $T\sim T^{*}$, $C(T)$ still displays a $T^2$ dependence with a correction:
\begin{equation}\label{Eq_SHeat_03}
C(T)\approx\frac{6N\zeta(3)\kappa T^2}{\pi\sqrt{\eta_x \eta_y }v^2_F}\bigg[1-\frac{1}{12\zeta(3)}\left(\frac{T^{*} }{T}\right)^2\bigg].
\end{equation}
On the other hand, when $T\ll T^*$, $C(T)$ displays an exponential suppression in $T$, as expected for gapped systems:
\begin{equation}\label{Eq_SHeat_04}
C(T)\approx \frac{N \kappa (T^{*})^2}{\pi \sqrt{\eta_x \eta_y } v^2_F } \left(\frac{T^{*}}{T}\right)  \exp{\left(-\frac{T^{*}}{T}\right)}
\end{equation}

Note that this exponential suppression is not as sharp as the one occurring in a conventional superconductor described by the BCS theory \cite{Bardeen-PR(1957)}, since $T^{*} \sim \sqrt{r} |\Delta|$ can still be significantly smaller than $|\Delta|$ close enough to the AFM-QCP.

The full dependence of the specific heat on both the temperature and the distance to the QCP $r$ is illustrated in Fig. \ref{Specific_Heat}(a). The main result is the $T^2$ dependence of $C(T)$ at the AFM-QCP. Importantly, even away from the QCP, there is a clear crossover from a $T^2$ dependence to an exponential suppression; the corresponding crossover temperature is $T^{*} \sim \sqrt{r} |\Delta|$ discussed above. Fig. \ref{Specific_Heat}(b) shows the corresponding enhancement of $C(T)/T^2$ for a fixed temperature as the QCP is approached.

We emphasize that the analysis presented here is restricted to the disordered-AFM side of the magnetic QCP, since we are not taking into account the effects of a finite staggered magnetization, but only the presence of AFM fluctuations inside a long-range ordered SC phase. A full treatment of the behavior of the specific heat as the system enters the SC-AFM coexistence phase will require detailed knowledge of the band structure and other microscopic details. This is beyond the scope of this work, where we instead focus on universal properties that do not depend crucially on microscopic considerations.

\begin{figure}[t]
\centering \includegraphics[width=0.85\linewidth]{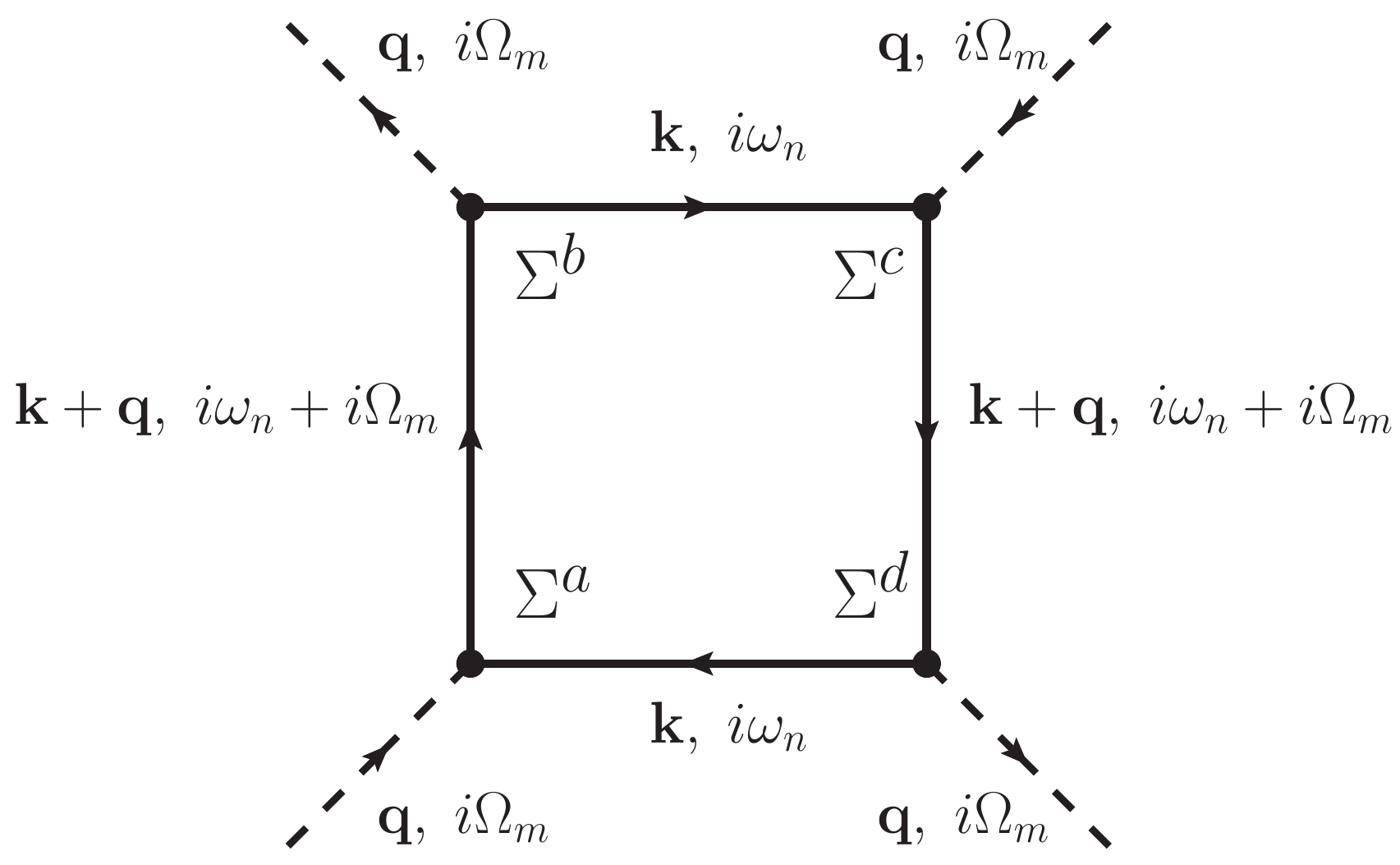}
\caption{Four-leg Feynman diagram used to determine the temperature dependence of the AFM mass on the disordered side of the AFM-QCP (see Eq. \eqref{AFM_mass_Eq_a}). The solid lines represent the fermionic propagators, whereas the dashed lines denote the AFM order parameter.}\label{Four_Legged_Diagram}
\end{figure}

Also, in our study we focused only on the magnetic contribution to the free energy and neglected the electronic part.  The conventional reasoning for this is that electronic states are gapped out and only contribute $e^{-|\Delta|/T}$  to the specific heat.  Near a QCP, one has to be a bit more careful because the fermionic self-energy
is singular at $r=0$ and at a finite $T$ due to a singular contribution from static spin fluctuations:  $\Sigma (\mathbf{k},i\omega_n) \propto T \chi_L G(\mathbf{k}+\mathbf{Q}, i\omega_n)$, where $\chi_L = \int d^2 q \chi (\mathbf{q},0)$. This singular self-energy gives rise to thermal precursors to the magnetic state by splitting the spectral function peak at $\omega = |\Delta| $ into two peaks at a smaller and a larger frequency.  For our purposes, the relevant question is whether this  $\Sigma (\mathbf{k}, i\omega_n)$  gives rise to a singular fermionic contribution to the free energy. To address this question, we analyze the fermionic part of the free energy
\begin{align}
& F_{el} = -T \sum_{\omega_n} \int_{\mathbf{k}} \log{\left[\epsilon^2_{\mathbf{k}} + (\omega_n +  \Sigma (\mathbf{k}, i\omega_n))^2 + \Phi^2 (\mathbf{k}, i\omega_n)\right]} \nonumber\\
& -2T \sum_{\omega_n} \int_{\mathbf{k}} \hspace{ -0.1 cm} \left[ - i \Sigma (\mathbf{k}, i\omega_n) G(\mathbf{k}, i\omega_n) + i \Phi (\mathbf{k}, i\omega_n) F(\mathbf{k}, i\omega_n)\right].
\label{F_el}
\end{align}
Here we explicitly introduced the anomalous self-energy $\Phi$ and the anomalous Green's function $F$.
The poles in $G$ and $F$ are at a finite $\omega \geq |\Delta|$ and $\Phi \approx |\Delta|$, so the non-exponential contribution can only come from the normal self-energy $\Sigma$.  However, expanding Eq. \eqref{F_el} in $\Sigma$, we find that at least the leading-order contribution cancels out, i.e., there is no non-exponential contribution from the fermionic part of the free energy. To check what happens for higher-order terms in $\Sigma$, one would need to include higher-order skeleton diagrams.

\section{Impact of the mode-mode coupling on the specific heat}\label{Sec_V}

We now analyze how the specific heat near the AFM-QCP is affected by Gaussian corrections from the $b(\delta M)^4$ term in the Ginzburg-Landau functional. Because fermions are gapped, the Ginzburg-Landau functional for soft AFM fluctuations is the soft-cutoff version of the effective action for the $O(N)$ nonlinear $\sigma$-model near a QCP. Therefore, the specific heat should have the same functional form as in the nonlinear $\sigma$ model-based analysis (see Ref. \cite{Chubukov-PRB(1994)}). In fact, the Gaussian approximation is equivalent to the $N \to \infty$ limit in the nonlinear $\sigma$-model analysis. Within this approximation, we obtain the following expression for the renormalized distance to the QCP, which we denote $m^2(T)$ hearafter ($m$ is also called the ``AFM mass"):
\begin{equation}\label{AFM_mass_Eq_a}
m^2(T) = r + b N T \sum_{\Omega_m} \int_\mathbf{q} \frac{ \nu^{-1}_0}{ m^2 (T) + \eta(\theta)(v_Fq)^2 + \kappa \Omega^2_m},
\end{equation}
where the factor $b$ in the $b(\delta M)^4$ term present in the effective action is the four-leg vertex shown in Fig. \ref{Four_Legged_Diagram}. One expects it to be of the order of $b \sim (\nu_0 g_{\text{afm}})^4/|\Delta|^2$. Indeed, we explicitly computed $b$ for perfect nesting (i.e., $\delta_0 = \delta_2 = 0$) and found $b  = 2(\nu_0 g_{\text{afm}})^4/|\Delta|^2$.

\begin{figure*}[t]
\centering \includegraphics[width=0.49\linewidth]{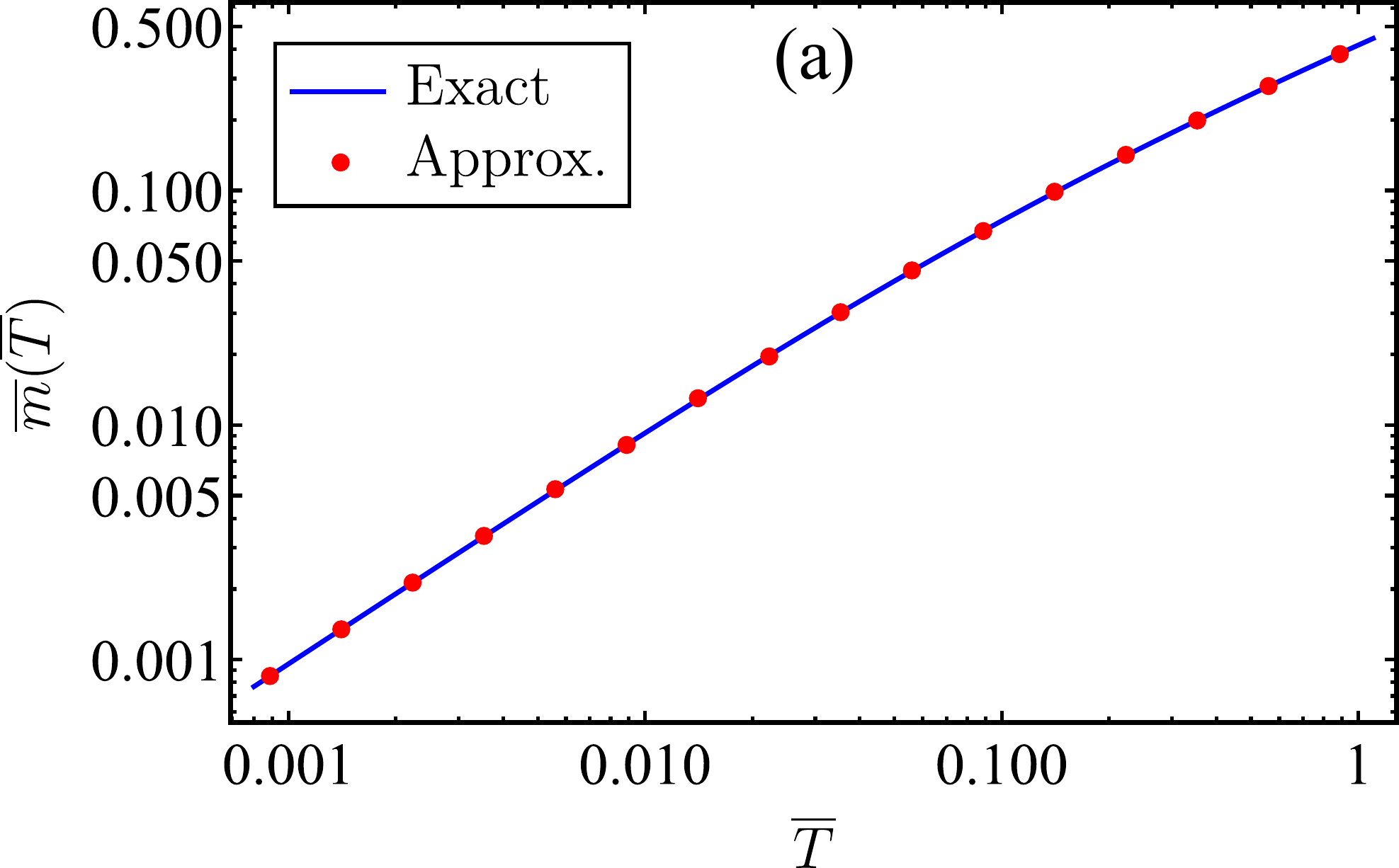}\hfill{}\includegraphics[width=0.49\linewidth]{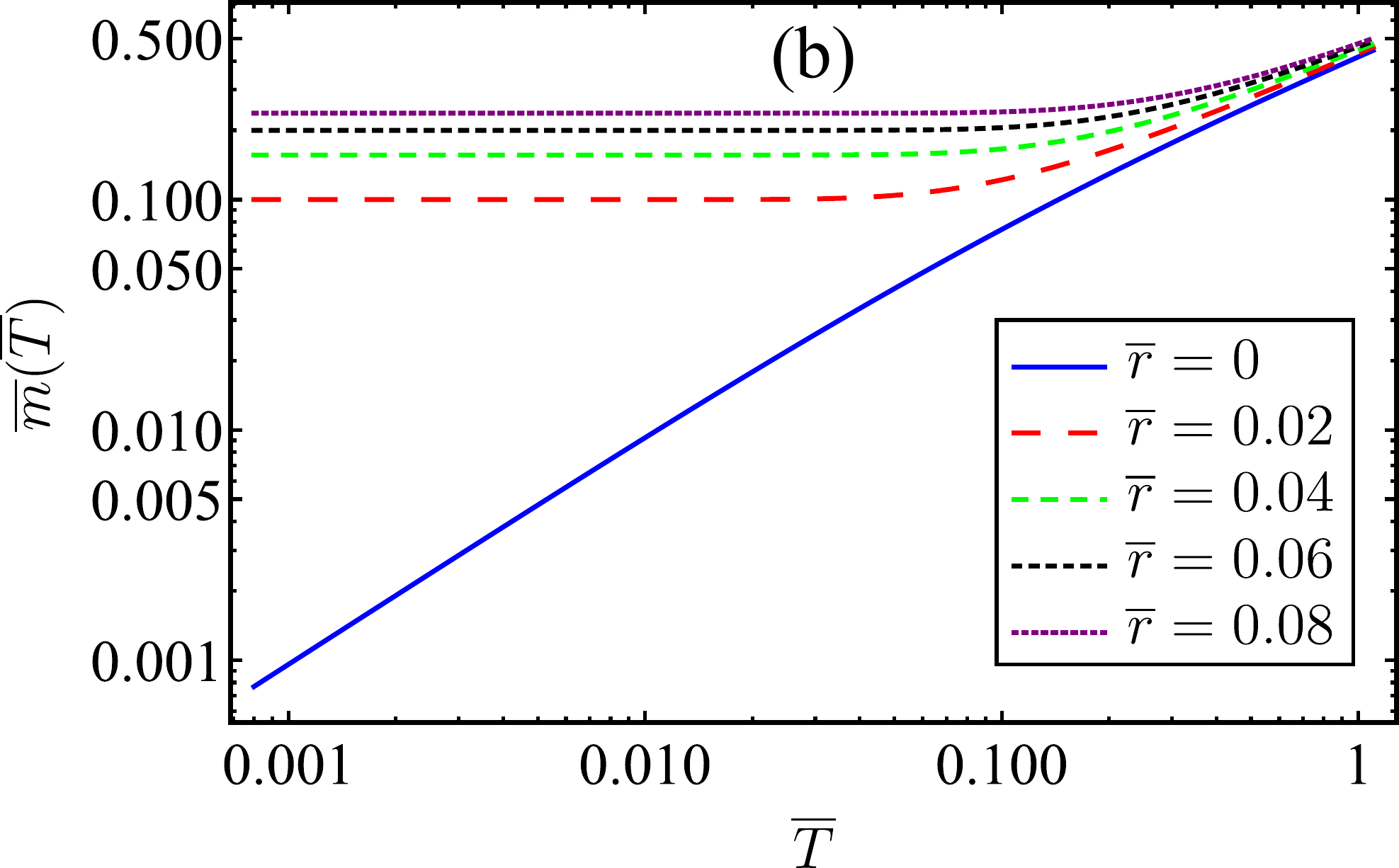}
\caption{(a) Temperature dependence of the AFM mass $\overline{m}(\overline{T})$ when the system is located at the QCP. The red dots here refer to the asymptotic behavior of $\overline{m}(\overline{T})$ obtained in Eq. \eqref{Approx_Mass_Sol}. Notice that it matches perfectly the exact solution for the AFM mass. (b) When the distance $r$ to the QCP becomes finite, $\overline{m}(\overline{T})$ approaches finite plateaus as the temperature decreases towards zero. These panels were obtained by setting $\overline{\lambda}=0.2$.}\label{AFM_Mass}
\end{figure*}

Regularizing the ultraviolet divergence on the right-hand side of Eq. \eqref{AFM_mass_Eq_a} by incorporating the contribution at $T = 0$ and $m(T = 0)=0$ into a re-definition of $r$, and then evaluating explicitly the frequency sum and the momentum integral, we obtain
\begin{equation}\label{AFM_mass_Eq}
m^2(T) = r + \lambda \sqrt{\kappa} T \log\bigg[\frac{1}{2}\operatorname{csch}\bigg(\frac{m(T)}{2\sqrt{\kappa}T}\bigg)\bigg],
\end{equation}
where $\lambda \equiv \dfrac{N (\nu_0 g_{\text{afm}})^{4}}{\pi \sqrt{\eta_x\eta_y \kappa} v^2_F\nu_0 |\Delta|^2}$ is a dimensionless coupling.
Substituting $\eta_{x,y} = \overline{\eta}_{x,y}/|\Delta|^2$, $\kappa = \overline{\kappa}/|\Delta|^2$, and $\pi v^2_F \nu_0 = E_F$, we obtain $\lambda=\dfrac{N (\nu_0 g_{\text{afm}})^{4}}{\sqrt{\overline{\eta}_x\overline{\eta}_y\overline{\kappa}}}\dfrac{|\Delta|}{E_F}$.  We assume that $|\Delta| \ll E_F$, hence $\lambda \ll 1$.
It is convenient to introduce $\overline{m} \equiv m/|\Delta|\sqrt{\kappa}$ and $\overline{r} \equiv r/|\Delta|^2 \kappa$, $\overline{T} \equiv T/|\Delta|$, $\overline{\lambda} \equiv \lambda/|\Delta| \sqrt{\kappa} \sim \lambda$,  and re-express Eq. \eqref{AFM_mass_Eq} as
\begin{equation}\label{Transc_AFM_Mass_Eq}
\overline{m}^2(\overline{T})=\overline{r} + \overline{\lambda}\; \overline{T}\log\bigg[\frac{1}{2}\operatorname{csch}\bigg(\frac{\overline{m}(\overline{T})}{2\overline{T}}\bigg)\bigg].
\end{equation}
 At $T=0$, Eq. \eqref{Transc_AFM_Mass_Eq} reduces to
\begin{equation}\label{Zero_Temp_Mass_01}
\overline{m}^2_0 + \frac{1}{2} \overline{\lambda} \overline{m}_0 = \overline{r},
\end{equation}
where $\overline{m}_0$ is the AFM zero-temperature mass. We see that at the smallest $\overline{m}_0$, the mode-mode coupling term becomes the dominant one as it contains a smaller power of $\overline{m}_0$.  Solving Eq. \eqref{Zero_Temp_Mass_01}, we obtain
\begin{equation}\label{Zero_Temp_Mass_02}
\overline{m}_0=\frac{\overline{\lambda}}{4}\bigg(\sqrt{1+\frac{16\overline{r}}{\overline{\lambda}^2}}-1\bigg).
\end{equation}
At  $\overline{r} \gg \overline{\lambda}^2$, the mode-mode coupling is irrelevant and $\overline{m}_0 \approx\sqrt{\overline{r}}$, i.e., $\overline{m}^2_0 \approx \overline{r}$. In the opposite limit $\overline{r} \ll \overline{\lambda}^2$, the dependence of $\overline{m}_0 $ on $\overline{r}$ is determined by the mode-mode coupling, and $\overline{m}_0\approx 2\overline{r}/\overline{\lambda}$.  The crossover between the two dependencies occurs at $\overline{r} \sim {\overline{\lambda}}^2$.

 The same happens when the system is at the AFM-QCP and one considers finite temperatures.  At $\overline{T} \ll \overline{\lambda}$, we have from Eq. \eqref{Transc_AFM_Mass_Eq}:
 \begin{equation}\label{Zero_Temp_Mass_03}
\overline{m} (\overline{T})  = \Theta\overline{T},
\end{equation}
where $\Theta \approx 0.962424$ is the solution of $ \operatorname{csch} (\Theta/2) =2$ \cite{Chubukov-PRB(1994)}. In the opposite limit $\overline{T} \gg \overline{\lambda}$, $\overline{m} (\overline{T})$ becomes much smaller than $\overline{T}$:
\begin{equation}\label{Zero_Temp_Mass_04}
\overline{m} (\overline{T})  \approx \overline{T} \sqrt{\frac{\overline{\lambda}}{\overline{T}} \log{\frac{\overline{T}}{\overline{\lambda}}}}.
\end{equation}
The  crossover between the two regimes occurs at $\overline{T} \sim \overline{\lambda}$, i.e., at
$T \sim \lambda/\sqrt{\kappa} \sim \lambda \Delta \ll \Delta$.

We solved Eq. \eqref{Transc_AFM_Mass_Eq} numerically and show the result in Fig. \ref{AFM_Mass}. At $r=0$, the behavior of $\overline{m} (\overline{T})$  can be reproduced,  to a surprisingly good accuracy, if we approximate the right-hand side of Eq. \eqref{Transc_AFM_Mass_Eq} by expanding to second order in $x=\overline{m} (\overline{T})/\overline{T}$. Namely,  if we approximate the logarithm in that equation as
\begin{equation}
\log\bigg[\frac{1}{2}\operatorname{csch}\bigg(\frac{x}{2}\bigg)\bigg]\approx\log\bigg(\frac{1}{x}\bigg)-\frac{x^2}{24},
\end{equation}
 we obtain
\begin{equation}\label{Approx_Mass_Sol}
\overline{m}(\overline{T})=\overline{T}\exp\bigg\lbrace-\frac{1}{2}
W\bigg[\frac{2\overline{T}}{\overline{\lambda}}\bigg(1+\frac{\overline{\lambda}}{24\overline{T}}\bigg)\bigg]\bigg\rbrace,
\end{equation}
where $W(x)$ is the so-called Lambert function. We plot this dependence in Fig. \ref{AFM_Mass}(a) along with the exact solution. We see that they are extremely close for all $\overline{T} = T/|\Delta|$.  At the smallest $T$, Eq. \eqref{Approx_Mass_Sol} yields $\overline{m} (\overline{T})  = \Theta^{*} \overline{T}$, where $\Theta^{*} \approx 0.962161$ is extremely close to the exact $\Theta \approx 0.962424$. In addition, we show in Fig. \ref{AFM_Mass}(b) that as the QCP distance $\overline{r}$ is increased, the AFM mass $\overline{m} (\overline{T})$ becomes more flat, which indicates less influence of mode-mode coupling.

\begin{figure}[t]
\centering \includegraphics[width=1.0\linewidth]{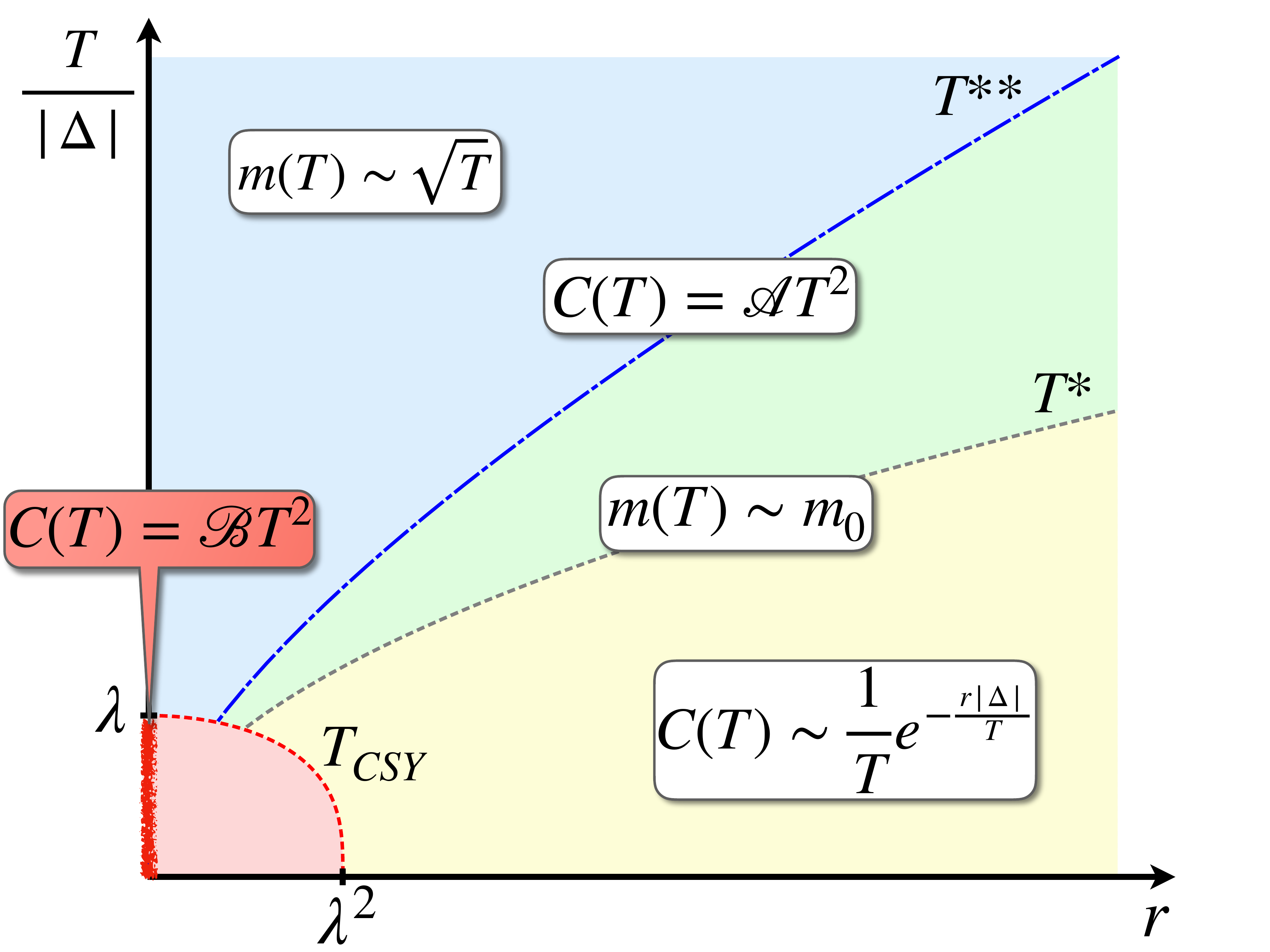}
\caption{Schematic phase diagram derived from the behavior of the AFM mass and the specific heat, when mode-mode coupling is taken into account. Here, all dashed lines denote crossover temperatures. In the region below $T^{**}$, one finds $m(T)\sim m_0$, where $m_0$ is the AFM mass at zero temperature, while above this temperature scale, $m(T)$ is approximately described by Eq. \eqref{Zero_Temp_Mass_04}. The specific heat behaves as $C(T)=\mathcal{A}T^2$ for temperatures larger than $T^{*}$, while it displays an exponential decay in the opposite limit. Besides, $T_\text{CSY}$ refers to the boundaries of a region described by the physics of the two-dimensional non-linear $\sigma$-model \cite{Chubukov-PRB(1994)}. Here, the temperature dependence of the specific heat also changes from $T^2$ to an exponentially suppressed behavior for $r>0$; at the QCP, it becomes $C(T)=\mathcal{B}T^2$ when $T \ll \lambda |\Delta|$. In addition, the zero-temperature AFM mass within this region behaves as $m_0\sim r/\lambda$.}\label{SHeat_Phases}
\end{figure}

Having determined the temperature dependence of the AFM mass due to mode-mode coupling, we now move on to investigate its effects on the behavior of the specific heat. Our main results are presented in the phase diagram displayed in Fig. \ref{SHeat_Phases}. At the AFM-QCP and for temperatures $T\ll\lambda |\Delta|$, the specific heat behaves asymptotically as
\begin{equation}\label{Eq_SHeat_05}
C_\text{QCP}(T)=\frac{6N\kappa}{\pi\sqrt{\eta_x\eta_y}v^2_F}\left(\Upsilon(\Theta^2)+\frac{\Theta^3}{6}\right)T^2,
\end{equation}
where, as defined before, $\Upsilon(z)=\sqrt{z}\operatorname{Li}_2(e^{-\sqrt{z}})+\operatorname{Li}_3(e^{-\sqrt{z}})$ and $\Upsilon(\Theta^2) + \Theta^{3}/6 = 4 \zeta (3)/5$ (see Ref. \cite{Sachdev-PLB(1993)}). The second term in parentheses comes from $F_0$, defined after Eq. \eqref{Eq_Free_Energy}, once we replace $r$ in $E_\mathbf{q}$ by $m^2 (T)$. The numerical factor $\Upsilon(\Theta^2) + \Theta^{3}/6 = 4 \zeta (3)/5$ in Eq. \eqref{Eq_SHeat_05} agrees with the result for the specific heat in Ref. \cite{Chubukov-PRB(1994)} (note that in Ref. \cite{Chubukov-PRB(1994)}  the free energy $F$ is defined with an extra $1/2$ compared to our Eq. \eqref{F_final}).

By comparing this term with Eq. \eqref{Eq_SHeat_02}, we see that the effect of mode-mode coupling here is to reduce the value of the specific-heat coefficient, i.e., $\zeta(3)$ is replaced by $\Upsilon(\Theta^2) + \Theta^{3}/6$. In addition, when the temperature evolves to $T\gg\lambda |\Delta|$, we still find that $C(T)$ depends on the temperature as $T^2$, because of the rapid decay of $m^2(T)/T^2$ obtained in that situation [see Eqs. \eqref{Zero_Temp_Mass_04} and \eqref{Approx_Mass_Sol}].

Away from the AFM-QCP and below a certain temperature scale $T^{**} \equiv T^{*}e^{W(\overline{r}/\overline{\lambda})}$ where the AFM mass is of order of $\overline{m}_0$, the leading contribution to the specific heat becomes
\begin{equation}
C(T)=\frac{6N\kappa T^2}{\pi\sqrt{\eta_x\eta_y}v^2_F}\mathcal{K}\bigg[\frac{m^2_0}{\kappa T^2}\bigg],
\end{equation}
where $m_0=\sqrt{\overline{\kappa}}\;\overline{m}_0$ is obtained from Eq. \eqref{Zero_Temp_Mass_02}. By using the asymptotic expressions for $m_0$ derived on both sides of the crossover point $r \sim \lambda^2$, we also find:
\begin{equation}
C(T) =
\begin{dcases}
\dfrac{6N\kappa T^2}{\pi\sqrt{\eta_x\eta_y}v^2_F}\mathcal{K}\left[\dfrac{4r^2 |\Delta|^2}{\overline{\kappa}\lambda^2 T^2}\right], & \text{if } r\ll \lambda^2, \\
\dfrac{6N\kappa T^2}{\pi\sqrt{\eta_x\eta_y}v^2_F}\mathcal{K}\left[\dfrac{r |\Delta|^2}{\overline{\kappa} T^2}\right], & \text{if } r \gg \lambda^2.
\end{dcases}
\end{equation}
The leading temperature dependence of $C(T)$ obtained here follows closely the results obtained in Eqs. \eqref{Eq_SHeat_03} and \eqref{Eq_SHeat_04} and is also indicated in Fig. \ref{SHeat_Phases}. The only difference when mode-mode coupling is taken into account is in the prefactor of $C(T)$ for the regime $r \ll \lambda^2$. This region is the zero-temperature projection of a crossover line $T_\text{CSY}$, in which the behavior of the system mimics that of a nonlinear $\sigma$-model describing the quantum-disordered regime of the two-dimensional Heisenberg model close to a QCP (Ref. \cite{Chubukov-PRB(1994)}). However, the area below $T_\text{CSY}$ in the phase diagram of our multi-band model is expected to be small, since it scale as $\lambda^3$ (see Fig. \ref{SHeat_Phases}).

Finally for temperatures larger than $T^{**}$, we can safely use the AFM mass obtained in Eq. \eqref{Zero_Temp_Mass_04} to evaluate the free energy and then the specific heat. As already emphasized, $m^2(T)/T^2$ decays rapidly as the temperature increases in that situation. As a result, we also obtain $C(T)\propto T^2$. The main distinction in the behavior of the specific heat for $T^{*}<T<T^{**}$ and $T>T^{**}$ is therefore the value of its coefficient.

\section{Conclusions and outlook}\label{Sec_VI}

In summary, we showed that the specific-heat of a fully gapped superconductor acquires a power-law temperature dependence at low-$T$ arising from the contributions of AFM fluctuations associated with a QCP inside the SC dome. Precisely at the QCP, a $T^2$ dependence persists down to zero temperature. Away from the QCP, this $T^2$ dependence changes at a crossover temperature $T^{*} \sim \sqrt{r} |\Delta|$ to an exponential suppression. Importantly, $T^{*}$ can be very small close enough to the QCP. The fact that the $T^2$ dependence persists over a certain temperature range even away from the QCP is important, because in some materials, such as Ba(Fe$_{1-x}$Co$_x$)$_2$As$_2$, the AFM transition line displays a back-bending once it crosses the SC dome \cite{Fernandes-PRB(2010)}, making the AFM-QCP inaccessible from the disordered phase just by lowering the temperature.

We also showed that the temperature variation of the AFM mass due to mode-mode coupling does not change crucially this behavior of the specific heat. However, we identified a region close to the AFM-QCP, in which the properties of the multi-band model analyzed here are equivalent to those of the nonlinear $\sigma$-model describing the quantum-disordered side of a two-dimensional Heisenberg model \cite{Chubukov-PRB(1994)}.

We argued that these findings are universal and do not dependent on microscopic considerations, as they follow from the form of the AFM propagator inside the SC state, which gives a non-analytic contribution to the free energy. These results provide an unambiguous method to detect an AFM-QCP inside the SC dome, as the non-trivial $T^2$ dependence is very different than the exponential $e^{-|\Delta|/T}$ behavior expected for a fully gapped superconductor without quantum critical AFM fluctuations. In this context, it would be interesting to experimentally revisit the phase diagram of known fully-gapped superconductors that coexist microscopically with AFM. The main candidates are the BaFe$_2$(As$_{1-x}$P$_x$)$_2$ and Ba(Fe$_{1-x}$Co$_x$)$_2$As$_2$ iron-based superconductors, for which London penetration depth measurements already suggest the existence of an AFM-QCP inside the dome \cite{Hashimoto-S(2012),Prozorov-NJP(2020)}. Another interesting candidate is stoichiometric CaKFe$_4$As$_4$, which has been proposed to be very close to a magnetic QCP \cite{Furukawa-PRL(2018)}, with no accompanying nematic order \cite{Canfield-NPJ(2018)}. Specific heat measurements would provide a clear thermodynamic signature for such an AFM-QCP, avoiding some of the theoretical issues arising from the interpretation of the penetration depth measurements. In addition, we also expect that other fully gapped superconductors in proximity to either a magnetic or non-magnetic QCP with dynamical exponent $z = 2$ will exhibit much of the specific-heat phenomenology described here.

\begin{acknowledgments}

We would like to thank Eduardo Miranda, Hermann Freire, and specially Subir Sachdev for stimulating discussions.
V.S.deC. acknowledges the financial support from FAPESP under Grants Nos.
2016/05069-7 and 2017/16911-3. A.V.C. was supported by the U.S. Department of Energy,
Office of Science, Basic Energy Sciences, under Award No. DE-SC0014402. R.M.F. was supported by the U.S. Department of Energy,
Office of Science, Basic Energy Sciences, under Award No. DE-SC0020045.

\end{acknowledgments}


\begin{thebibliography}{51}%
\makeatletter
\providecommand \@ifxundefined [1]{%
 \@ifx{#1\undefined}
}%
\providecommand \@ifnum [1]{%
 \ifnum #1\expandafter \@firstoftwo
 \else \expandafter \@secondoftwo
 \fi
}%
\providecommand \@ifx [1]{%
 \ifx #1\expandafter \@firstoftwo
 \else \expandafter \@secondoftwo
 \fi
}%
\providecommand \natexlab [1]{#1}%
\providecommand \enquote  [1]{``#1''}%
\providecommand \bibnamefont  [1]{#1}%
\providecommand \bibfnamefont [1]{#1}%
\providecommand \citenamefont [1]{#1}%
\providecommand \href@noop [0]{\@secondoftwo}%
\providecommand \href [0]{\begingroup \@sanitize@url \@href}%
\providecommand \@href[1]{\@@startlink{#1}\@@href}%
\providecommand \@@href[1]{\endgroup#1\@@endlink}%
\providecommand \@sanitize@url [0]{\catcode `\\12\catcode `\$12\catcode
  `\&12\catcode `\#12\catcode `\^12\catcode `\_12\catcode `\%12\relax}%
\providecommand \@@startlink[1]{}%
\providecommand \@@endlink[0]{}%
\providecommand \url  [0]{\begingroup\@sanitize@url \@url }%
\providecommand \@url [1]{\endgroup\@href {#1}{\urlprefix }}%
\providecommand \urlprefix  [0]{URL }%
\providecommand \Eprint [0]{\href }%
\providecommand \doibase [0]{http://dx.doi.org/}%
\providecommand \selectlanguage [0]{\@gobble}%
\providecommand \bibinfo  [0]{\@secondoftwo}%
\providecommand \bibfield  [0]{\@secondoftwo}%
\providecommand \translation [1]{[#1]}%
\providecommand \BibitemOpen [0]{}%
\providecommand \bibitemStop [0]{}%
\providecommand \bibitemNoStop [0]{.\EOS\space}%
\providecommand \EOS [0]{\spacefactor3000\relax}%
\providecommand \BibitemShut  [1]{\csname bibitem#1\endcsname}%
\let\auto@bib@innerbib\@empty
\bibitem [{\citenamefont {Sachdev}\ and\ \citenamefont
  {Keimer}(2011)}]{Sachdev-PT(2011)}%
  \BibitemOpen
  \bibfield  {author} {\bibinfo {author} {\bibfnamefont {S.}~\bibnamefont
  {Sachdev}}\ and\ \bibinfo {author} {\bibfnamefont {B.}~\bibnamefont
  {Keimer}},\ }\bibfield  {title} {\bibinfo {title} {{Quantum criticality}},\
  }\href {\doibase 10.1063/1.3554314} {\bibfield  {journal} {\bibinfo
  {journal} {Phys. Today}\ }\textbf {\bibinfo {volume} {64}},\ \bibinfo {pages}
  {29} (\bibinfo {year} {2011})}\BibitemShut {NoStop}%
\bibitem [{\citenamefont {Vojta}(2003)}]{Vojta-RPP(2003)}%
  \BibitemOpen
  \bibfield  {author} {\bibinfo {author} {\bibfnamefont {M.}~\bibnamefont
  {Vojta}},\ }\bibfield  {title} {\bibinfo {title} {{Quantum phase
  transitions}},\ }\href {\doibase 10.1088/0034-4885/66/12/r01} {\bibfield
  {journal} {\bibinfo  {journal} {Rep. Prog. Phys.}\ }\textbf {\bibinfo
  {volume} {66}},\ \bibinfo {pages} {2069} (\bibinfo {year}
  {2003})}\BibitemShut {NoStop}%
\bibitem [{\citenamefont {L\"ohneysen}\ \emph {et~al.}(2007)\citenamefont
  {L\"ohneysen}, \citenamefont {Rosch}, \citenamefont {Vojta},\ and\
  \citenamefont {W\"olfle}}]{Lohneysen-RMP(2007)}%
  \BibitemOpen
  \bibfield  {author} {\bibinfo {author} {\bibfnamefont {H.~v.}\ \bibnamefont
  {L\"ohneysen}}, \bibinfo {author} {\bibfnamefont {A.}~\bibnamefont {Rosch}},
  \bibinfo {author} {\bibfnamefont {M.}~\bibnamefont {Vojta}}, \ and\ \bibinfo
  {author} {\bibfnamefont {P.}~\bibnamefont {W\"olfle}},\ }\bibfield  {title}
  {\bibinfo {title} {{Fermi-liquid instabilities at magnetic quantum phase
  transitions}},\ }\href {\doibase 10.1103/RevModPhys.79.1015} {\bibfield
  {journal} {\bibinfo  {journal} {Rev. Mod. Phys.}\ }\textbf {\bibinfo {volume}
  {79}},\ \bibinfo {pages} {1015} (\bibinfo {year} {2007})}\BibitemShut
  {NoStop}%
\bibitem [{\citenamefont {Sachdev}(2011)}]{Sacdev-CUP(2011)}%
  \BibitemOpen
  \bibfield  {author} {\bibinfo {author} {\bibfnamefont {S.}~\bibnamefont
  {Sachdev}},\ }\href@noop {} {\emph {\bibinfo {title} {{Quantum Phase
  Transitions}}}},\ \bibinfo {edition} {2nd}\ ed.\ (\bibinfo  {publisher}
  {Cambridge University Press},\ \bibinfo {address} {Cambridge},\ \bibinfo
  {year} {2011})\BibitemShut {NoStop}%
\bibitem [{\citenamefont {Scalapino}(2012)}]{Scalapino2012}%
  \BibitemOpen
  \bibfield  {author} {\bibinfo {author} {\bibfnamefont {D.~J.}\ \bibnamefont
  {Scalapino}},\ }\bibfield  {title} {\bibinfo {title} {{A common thread: The
  pairing interaction for unconventional superconductors}},\ }\href {\doibase
  10.1103/RevModPhys.84.1383} {\bibfield  {journal} {\bibinfo  {journal} {Rev.
  Mod. Phys.}\ }\textbf {\bibinfo {volume} {84}},\ \bibinfo {pages} {1383}
  (\bibinfo {year} {2012})}\BibitemShut {NoStop}%
\bibitem [{\citenamefont {Abanov}\ \emph {et~al.}(2003)\citenamefont {Abanov},
  \citenamefont {Chubukov},\ and\ \citenamefont {Schmalian}}]{Abanov-AP(2003)}%
  \BibitemOpen
  \bibfield  {author} {\bibinfo {author} {\bibfnamefont {A.}~\bibnamefont
  {Abanov}}, \bibinfo {author} {\bibfnamefont {A.~V.}\ \bibnamefont
  {Chubukov}}, \ and\ \bibinfo {author} {\bibfnamefont {J.}~\bibnamefont
  {Schmalian}},\ }\bibfield  {title} {\bibinfo {title} {{Quantum-critical
  theory of the spin-fermion model and its application to cuprates: Normal
  state analysis}},\ }\href {\doibase 10.1080/0001873021000057123} {\bibfield
  {journal} {\bibinfo  {journal} {Adv. Phys.}\ }\textbf {\bibinfo {volume}
  {52}},\ \bibinfo {pages} {119} (\bibinfo {year} {2003})}\BibitemShut
  {NoStop}%
\bibitem [{\citenamefont {Metlitski}\ and\ \citenamefont
  {Sachdev}(2010)}]{Metlitski-PRBb(2010)}%
  \BibitemOpen
  \bibfield  {author} {\bibinfo {author} {\bibfnamefont {M.~A.}\ \bibnamefont
  {Metlitski}}\ and\ \bibinfo {author} {\bibfnamefont {S.}~\bibnamefont
  {Sachdev}},\ }\bibfield  {title} {\bibinfo {title} {{Quantum phase
  transitions of metals in two spatial dimensions. II. Spin density wave
  order}},\ }\href {\doibase 10.1103/PhysRevB.82.075128} {\bibfield  {journal}
  {\bibinfo  {journal} {Phys. Rev. B}\ }\textbf {\bibinfo {volume} {82}},\
  \bibinfo {pages} {075128} (\bibinfo {year} {2010})}\BibitemShut {NoStop}%
\bibitem [{\citenamefont {Proust}\ and\ \citenamefont
  {Taillefer}(2019)}]{Taillefer2019}%
  \BibitemOpen
  \bibfield  {author} {\bibinfo {author} {\bibfnamefont {C.}~\bibnamefont
  {Proust}}\ and\ \bibinfo {author} {\bibfnamefont {L.}~\bibnamefont
  {Taillefer}},\ }\bibfield  {title} {\bibinfo {title} {{The Remarkable
  Underlying Ground States of Cuprate Superconductors}},\ }\href {\doibase
  10.1146/annurev-conmatphys-031218-013210} {\bibfield  {journal} {\bibinfo
  {journal} {Annu. Rev. Condens. Matter Phys.}\ }\textbf {\bibinfo {volume}
  {10}},\ \bibinfo {pages} {409} (\bibinfo {year} {2019})}\BibitemShut
  {NoStop}%
\bibitem [{\citenamefont {Hayes}\ \emph {et~al.}(2016)\citenamefont {Hayes},
  \citenamefont {McDonald}, \citenamefont {Breznay}, \citenamefont {Helm},
  \citenamefont {Moll}, \citenamefont {Wartenbe}, \citenamefont {Shekhter},\
  and\ \citenamefont {Analytis}}]{Analytis2016}%
  \BibitemOpen
  \bibfield  {author} {\bibinfo {author} {\bibfnamefont {I.~M.}\ \bibnamefont
  {Hayes}}, \bibinfo {author} {\bibfnamefont {R.~D.}\ \bibnamefont {McDonald}},
  \bibinfo {author} {\bibfnamefont {N.~P.}\ \bibnamefont {Breznay}}, \bibinfo
  {author} {\bibfnamefont {T.}~\bibnamefont {Helm}}, \bibinfo {author}
  {\bibfnamefont {P.~J.}\ \bibnamefont {Moll}}, \bibinfo {author}
  {\bibfnamefont {M.}~\bibnamefont {Wartenbe}}, \bibinfo {author}
  {\bibfnamefont {A.}~\bibnamefont {Shekhter}}, \ and\ \bibinfo {author}
  {\bibfnamefont {J.~G.}\ \bibnamefont {Analytis}},\ }\bibfield  {title}
  {\bibinfo {title} {{Scaling between magnetic field and temperature in the
  high-temperature superconductor BaFe$_2$(As$_{1- x}$P$_x$)$_2$}},\ }\href
  {\doibase 10.1038/nphys3773} {\bibfield  {journal} {\bibinfo  {journal} {Nat.
  Phys.}\ }\textbf {\bibinfo {volume} {12}},\ \bibinfo {pages} {916} (\bibinfo
  {year} {2016})}\BibitemShut {NoStop}%
\bibitem [{\citenamefont {Fernandes}\ and\ \citenamefont
  {Schmalian}(2010)}]{Fernandes-PRB(2010)}%
  \BibitemOpen
  \bibfield  {author} {\bibinfo {author} {\bibfnamefont {R.~M.}\ \bibnamefont
  {Fernandes}}\ and\ \bibinfo {author} {\bibfnamefont {J.}~\bibnamefont
  {Schmalian}},\ }\bibfield  {title} {\bibinfo {title} {{Competing order and
  nature of the pairing state in the iron pnictides}},\ }\href {\doibase
  10.1103/PhysRevB.82.014521} {\bibfield  {journal} {\bibinfo  {journal} {Phys.
  Rev. B}\ }\textbf {\bibinfo {volume} {82}},\ \bibinfo {pages} {014521}
  (\bibinfo {year} {2010})}\BibitemShut {NoStop}%
\bibitem [{\citenamefont {Vorontsov}\ \emph {et~al.}(2010)\citenamefont
  {Vorontsov}, \citenamefont {Vavilov},\ and\ \citenamefont
  {Chubukov}}]{Vorontsov2010}%
  \BibitemOpen
  \bibfield  {author} {\bibinfo {author} {\bibfnamefont {A.~B.}\ \bibnamefont
  {Vorontsov}}, \bibinfo {author} {\bibfnamefont {M.~G.}\ \bibnamefont
  {Vavilov}}, \ and\ \bibinfo {author} {\bibfnamefont {A.~V.}\ \bibnamefont
  {Chubukov}},\ }\bibfield  {title} {\bibinfo {title} {{Superconductivity and
  spin-density waves in multiband metals}},\ }\href {\doibase
  10.1103/PhysRevB.81.174538} {\bibfield  {journal} {\bibinfo  {journal} {Phys.
  Rev. B}\ }\textbf {\bibinfo {volume} {81}},\ \bibinfo {pages} {174538}
  (\bibinfo {year} {2010})}\BibitemShut {NoStop}%
\bibitem [{\citenamefont {Fernandes}\ \emph {et~al.}(2013)\citenamefont
  {Fernandes}, \citenamefont {Maiti}, \citenamefont {W\"olfle},\ and\
  \citenamefont {Chubukov}}]{Fernandes2013}%
  \BibitemOpen
  \bibfield  {author} {\bibinfo {author} {\bibfnamefont {R.~M.}\ \bibnamefont
  {Fernandes}}, \bibinfo {author} {\bibfnamefont {S.}~\bibnamefont {Maiti}},
  \bibinfo {author} {\bibfnamefont {P.}~\bibnamefont {W\"olfle}}, \ and\
  \bibinfo {author} {\bibfnamefont {A.~V.}\ \bibnamefont {Chubukov}},\
  }\bibfield  {title} {\bibinfo {title} {{How Many Quantum Phase Transitions
  Exist Inside the Superconducting Dome of the Iron Pnictides?}}\ }\href
  {\doibase 10.1103/PhysRevLett.111.057001} {\bibfield  {journal} {\bibinfo
  {journal} {Phys. Rev. Lett.}\ }\textbf {\bibinfo {volume} {111}},\ \bibinfo
  {pages} {057001} (\bibinfo {year} {2013})}\BibitemShut {NoStop}%
\bibitem [{\citenamefont {R\o{}mer}\ \emph {et~al.}(2016)\citenamefont
  {R\o{}mer}, \citenamefont {Eremin}, \citenamefont {Hirschfeld},\ and\
  \citenamefont {Andersen}}]{Andersen2016}%
  \BibitemOpen
  \bibfield  {author} {\bibinfo {author} {\bibfnamefont {A.~T.}\ \bibnamefont
  {R\o{}mer}}, \bibinfo {author} {\bibfnamefont {I.}~\bibnamefont {Eremin}},
  \bibinfo {author} {\bibfnamefont {P.~J.}\ \bibnamefont {Hirschfeld}}, \ and\
  \bibinfo {author} {\bibfnamefont {B.~M.}\ \bibnamefont {Andersen}},\
  }\bibfield  {title} {\bibinfo {title} {{Superconducting phase diagram of
  itinerant antiferromagnets}},\ }\href {\doibase 10.1103/PhysRevB.93.174519}
  {\bibfield  {journal} {\bibinfo  {journal} {Phys. Rev. B}\ }\textbf {\bibinfo
  {volume} {93}},\ \bibinfo {pages} {174519} (\bibinfo {year}
  {2016})}\BibitemShut {NoStop}%
\bibitem [{\citenamefont {Silva}\ \emph {et~al.}(2018)\citenamefont {Silva},
  \citenamefont {Continentino},\ and\ \citenamefont {Barci}}]{Barci2018}%
  \BibitemOpen
  \bibfield  {author} {\bibinfo {author} {\bibfnamefont {N.~L.}\ \bibnamefont
  {Silva}}, \bibinfo {author} {\bibfnamefont {M.~A.}\ \bibnamefont
  {Continentino}}, \ and\ \bibinfo {author} {\bibfnamefont {D.~G.}\
  \bibnamefont {Barci}},\ }\bibfield  {title} {\bibinfo {title} {{Quantum
  corrections for the phase diagram of systems with competing order}},\ }\href
  {\doibase 10.1088/1361-648x/aac062} {\bibfield  {journal} {\bibinfo
  {journal} {J. Phys. Condens. Matter}\ }\textbf {\bibinfo {volume} {30}},\
  \bibinfo {pages} {225402} (\bibinfo {year} {2018})}\BibitemShut {NoStop}%
\bibitem [{\citenamefont {Foley}\ \emph {et~al.}(2019)\citenamefont {Foley},
  \citenamefont {Verret}, \citenamefont {Tremblay},\ and\ \citenamefont
  {S\'en\'echal}}]{Senechal2019}%
  \BibitemOpen
  \bibfield  {author} {\bibinfo {author} {\bibfnamefont {A.}~\bibnamefont
  {Foley}}, \bibinfo {author} {\bibfnamefont {S.}~\bibnamefont {Verret}},
  \bibinfo {author} {\bibfnamefont {A.-M.~S.}\ \bibnamefont {Tremblay}}, \ and\
  \bibinfo {author} {\bibfnamefont {D.}~\bibnamefont {S\'en\'echal}},\
  }\bibfield  {title} {\bibinfo {title} {{Coexistence of superconductivity and
  antiferromagnetism in the Hubbard model for cuprates}},\ }\href {\doibase
  10.1103/PhysRevB.99.184510} {\bibfield  {journal} {\bibinfo  {journal} {Phys.
  Rev. B}\ }\textbf {\bibinfo {volume} {99}},\ \bibinfo {pages} {184510}
  (\bibinfo {year} {2019})}\BibitemShut {NoStop}%
\bibitem [{\citenamefont {Julien}\ \emph {et~al.}(2009)\citenamefont {Julien},
  \citenamefont {Mayaffre}, \citenamefont {Horvati{\'{c}}}, \citenamefont
  {Berthier}, \citenamefont {Zhang}, \citenamefont {Wu}, \citenamefont {Chen},
  \citenamefont {Wang},\ and\ \citenamefont {Luo}}]{Julien2009}%
  \BibitemOpen
  \bibfield  {author} {\bibinfo {author} {\bibfnamefont {M.-H.}\ \bibnamefont
  {Julien}}, \bibinfo {author} {\bibfnamefont {H.}~\bibnamefont {Mayaffre}},
  \bibinfo {author} {\bibfnamefont {M.}~\bibnamefont {Horvati{\'{c}}}},
  \bibinfo {author} {\bibfnamefont {C.}~\bibnamefont {Berthier}}, \bibinfo
  {author} {\bibfnamefont {X.~D.}\ \bibnamefont {Zhang}}, \bibinfo {author}
  {\bibfnamefont {W.}~\bibnamefont {Wu}}, \bibinfo {author} {\bibfnamefont
  {G.~F.}\ \bibnamefont {Chen}}, \bibinfo {author} {\bibfnamefont {N.~L.}\
  \bibnamefont {Wang}}, \ and\ \bibinfo {author} {\bibfnamefont {J.~L.}\
  \bibnamefont {Luo}},\ }\bibfield  {title} {\bibinfo {title} {{Homogeneous vs.
  inhomogeneous coexistence of magnetic order and superconductivity probed by
  {NMR} in Co- and K-doped iron pnictides}},\ }\href {\doibase
  10.1209/0295-5075/87/37001} {\bibfield  {journal} {\bibinfo  {journal} {EPL
  (Europhysics Letters)}\ }\textbf {\bibinfo {volume} {87}},\ \bibinfo {pages}
  {37001} (\bibinfo {year} {2009})}\BibitemShut {NoStop}%
\bibitem [{\citenamefont {Marsik}\ \emph {et~al.}(2010)\citenamefont {Marsik},
  \citenamefont {Kim}, \citenamefont {Dubroka}, \citenamefont {R\"ossle},
  \citenamefont {Malik}, \citenamefont {Schulz}, \citenamefont {Wang},
  \citenamefont {Niedermayer}, \citenamefont {Drew}, \citenamefont {Willis},
  \citenamefont {Wolf},\ and\ \citenamefont {Bernhard}}]{Bernhard2010}%
  \BibitemOpen
  \bibfield  {author} {\bibinfo {author} {\bibfnamefont {P.}~\bibnamefont
  {Marsik}}, \bibinfo {author} {\bibfnamefont {K.~W.}\ \bibnamefont {Kim}},
  \bibinfo {author} {\bibfnamefont {A.}~\bibnamefont {Dubroka}}, \bibinfo
  {author} {\bibfnamefont {M.}~\bibnamefont {R\"ossle}}, \bibinfo {author}
  {\bibfnamefont {V.~K.}\ \bibnamefont {Malik}}, \bibinfo {author}
  {\bibfnamefont {L.}~\bibnamefont {Schulz}}, \bibinfo {author} {\bibfnamefont
  {C.~N.}\ \bibnamefont {Wang}}, \bibinfo {author} {\bibfnamefont
  {C.}~\bibnamefont {Niedermayer}}, \bibinfo {author} {\bibfnamefont {A.~J.}\
  \bibnamefont {Drew}}, \bibinfo {author} {\bibfnamefont {M.}~\bibnamefont
  {Willis}}, \bibinfo {author} {\bibfnamefont {T.}~\bibnamefont {Wolf}}, \ and\
  \bibinfo {author} {\bibfnamefont {C.}~\bibnamefont {Bernhard}},\ }\bibfield
  {title} {\bibinfo {title} {{Coexistence and Competition of Magnetism and
  Superconductivity on the Nanometer Scale in Underdoped
  ${\mathrm{BaFe}}_{1.89}{\mathrm{Co}}_{0.11}{\mathrm{As}}_{2}$}},\ }\href
  {\doibase 10.1103/PhysRevLett.105.057001} {\bibfield  {journal} {\bibinfo
  {journal} {Phys. Rev. Lett.}\ }\textbf {\bibinfo {volume} {105}},\ \bibinfo
  {pages} {057001} (\bibinfo {year} {2010})}\BibitemShut {NoStop}%
\bibitem [{\citenamefont {Wiesenmayer}\ \emph {et~al.}(2011)\citenamefont
  {Wiesenmayer}, \citenamefont {Luetkens}, \citenamefont {Pascua},
  \citenamefont {Khasanov}, \citenamefont {Amato}, \citenamefont {Potts},
  \citenamefont {Banusch}, \citenamefont {Klauss},\ and\ \citenamefont
  {Johrendt}}]{Johrendt2011}%
  \BibitemOpen
  \bibfield  {author} {\bibinfo {author} {\bibfnamefont {E.}~\bibnamefont
  {Wiesenmayer}}, \bibinfo {author} {\bibfnamefont {H.}~\bibnamefont
  {Luetkens}}, \bibinfo {author} {\bibfnamefont {G.}~\bibnamefont {Pascua}},
  \bibinfo {author} {\bibfnamefont {R.}~\bibnamefont {Khasanov}}, \bibinfo
  {author} {\bibfnamefont {A.}~\bibnamefont {Amato}}, \bibinfo {author}
  {\bibfnamefont {H.}~\bibnamefont {Potts}}, \bibinfo {author} {\bibfnamefont
  {B.}~\bibnamefont {Banusch}}, \bibinfo {author} {\bibfnamefont {H.-H.}\
  \bibnamefont {Klauss}}, \ and\ \bibinfo {author} {\bibfnamefont
  {D.}~\bibnamefont {Johrendt}},\ }\bibfield  {title} {\bibinfo {title}
  {{Microscopic Coexistence of Superconductivity and Magnetism in
  ${\mathrm{Ba}}_{1\ensuremath{-}x}{\mathrm{K}}_{x}{\mathrm{Fe}}_{2}{\mathrm{As}}_{2}$}},\
  }\href {\doibase 10.1103/PhysRevLett.107.237001} {\bibfield  {journal}
  {\bibinfo  {journal} {Phys. Rev. Lett.}\ }\textbf {\bibinfo {volume} {107}},\
  \bibinfo {pages} {237001} (\bibinfo {year} {2011})}\BibitemShut {NoStop}%
\bibitem [{\citenamefont {Ma}\ \emph {et~al.}(2012)\citenamefont {Ma},
  \citenamefont {Ji}, \citenamefont {Dai}, \citenamefont {Lu}, \citenamefont
  {Eom}, \citenamefont {Kim}, \citenamefont {Normand},\ and\ \citenamefont
  {Yu}}]{Ma2012}%
  \BibitemOpen
  \bibfield  {author} {\bibinfo {author} {\bibfnamefont {L.}~\bibnamefont
  {Ma}}, \bibinfo {author} {\bibfnamefont {G.~F.}\ \bibnamefont {Ji}}, \bibinfo
  {author} {\bibfnamefont {J.}~\bibnamefont {Dai}}, \bibinfo {author}
  {\bibfnamefont {X.~R.}\ \bibnamefont {Lu}}, \bibinfo {author} {\bibfnamefont
  {M.~J.}\ \bibnamefont {Eom}}, \bibinfo {author} {\bibfnamefont {J.~S.}\
  \bibnamefont {Kim}}, \bibinfo {author} {\bibfnamefont {B.}~\bibnamefont
  {Normand}}, \ and\ \bibinfo {author} {\bibfnamefont {W.}~\bibnamefont {Yu}},\
  }\bibfield  {title} {\bibinfo {title} {{Microscopic Coexistence of
  Superconductivity and Antiferromagnetism in Underdoped
  $\mathrm{Ba}({\mathrm{Fe}}_{1\mathbf{\ensuremath{-}}x}{\mathrm{Ru}}_{x}{)}_{2}{\mathrm{As}}_{2}$}},\
  }\href {\doibase 10.1103/PhysRevLett.109.197002} {\bibfield  {journal}
  {\bibinfo  {journal} {Phys. Rev. Lett.}\ }\textbf {\bibinfo {volume} {109}},\
  \bibinfo {pages} {197002} (\bibinfo {year} {2012})}\BibitemShut {NoStop}%
\bibitem [{\citenamefont {Cai}\ \emph {et~al.}(2013)\citenamefont {Cai},
  \citenamefont {Zhou}, \citenamefont {Ruan}, \citenamefont {Wang},
  \citenamefont {Chen}, \citenamefont {Lee},\ and\ \citenamefont
  {Wang}}]{Yayu2013}%
  \BibitemOpen
  \bibfield  {author} {\bibinfo {author} {\bibfnamefont {P.}~\bibnamefont
  {Cai}}, \bibinfo {author} {\bibfnamefont {X.}~\bibnamefont {Zhou}}, \bibinfo
  {author} {\bibfnamefont {W.}~\bibnamefont {Ruan}}, \bibinfo {author}
  {\bibfnamefont {A.}~\bibnamefont {Wang}}, \bibinfo {author} {\bibfnamefont
  {X.}~\bibnamefont {Chen}}, \bibinfo {author} {\bibfnamefont {D.-H.}\
  \bibnamefont {Lee}}, \ and\ \bibinfo {author} {\bibfnamefont
  {Y.}~\bibnamefont {Wang}},\ }\bibfield  {title} {\bibinfo {title}
  {{Visualizing the microscopic coexistence of spin density wave and
  superconductivity in underdoped NaFe$_{1- x}$Co$_x$As}},\ }\href {\doibase
  10.1038/ncomms2592} {\bibfield  {journal} {\bibinfo  {journal} {Nat.
  Commun.}\ }\textbf {\bibinfo {volume} {4}},\ \bibinfo {pages} {1} (\bibinfo
  {year} {2013})}\BibitemShut {NoStop}%
\bibitem [{\citenamefont {Cheung}\ \emph {et~al.}(2018)\citenamefont {Cheung},
  \citenamefont {Guguchia}, \citenamefont {Frandsen}, \citenamefont {Gong},
  \citenamefont {Yamakawa}, \citenamefont {Almeida}, \citenamefont {Onuorah},
  \citenamefont {Bonf\'a}, \citenamefont {Miranda}, \citenamefont {Wang},
  \citenamefont {Tam}, \citenamefont {Song}, \citenamefont {Cao}, \citenamefont
  {Cai}, \citenamefont {Hallas}, \citenamefont {Wilson}, \citenamefont
  {Munsie}, \citenamefont {Luke}, \citenamefont {Chen}, \citenamefont {Dai},
  \citenamefont {Jin}, \citenamefont {Guo}, \citenamefont {Ning}, \citenamefont
  {Fernandes}, \citenamefont {De~Renzi}, \citenamefont {Dai},\ and\
  \citenamefont {Uemura}}]{Sky2018}%
  \BibitemOpen
  \bibfield  {author} {\bibinfo {author} {\bibfnamefont {S.~C.}\ \bibnamefont
  {Cheung}}, \bibinfo {author} {\bibfnamefont {Z.}~\bibnamefont {Guguchia}},
  \bibinfo {author} {\bibfnamefont {B.~A.}\ \bibnamefont {Frandsen}}, \bibinfo
  {author} {\bibfnamefont {Z.}~\bibnamefont {Gong}}, \bibinfo {author}
  {\bibfnamefont {K.}~\bibnamefont {Yamakawa}}, \bibinfo {author}
  {\bibfnamefont {D.~E.}\ \bibnamefont {Almeida}}, \bibinfo {author}
  {\bibfnamefont {I.~J.}\ \bibnamefont {Onuorah}}, \bibinfo {author}
  {\bibfnamefont {P.}~\bibnamefont {Bonf\'a}}, \bibinfo {author} {\bibfnamefont
  {E.}~\bibnamefont {Miranda}}, \bibinfo {author} {\bibfnamefont
  {W.}~\bibnamefont {Wang}}, \bibinfo {author} {\bibfnamefont {D.~W.}\
  \bibnamefont {Tam}}, \bibinfo {author} {\bibfnamefont {Y.}~\bibnamefont
  {Song}}, \bibinfo {author} {\bibfnamefont {C.}~\bibnamefont {Cao}}, \bibinfo
  {author} {\bibfnamefont {Y.}~\bibnamefont {Cai}}, \bibinfo {author}
  {\bibfnamefont {A.~M.}\ \bibnamefont {Hallas}}, \bibinfo {author}
  {\bibfnamefont {M.~N.}\ \bibnamefont {Wilson}}, \bibinfo {author}
  {\bibfnamefont {T.~J.~S.}\ \bibnamefont {Munsie}}, \bibinfo {author}
  {\bibfnamefont {G.}~\bibnamefont {Luke}}, \bibinfo {author} {\bibfnamefont
  {B.}~\bibnamefont {Chen}}, \bibinfo {author} {\bibfnamefont {G.}~\bibnamefont
  {Dai}}, \bibinfo {author} {\bibfnamefont {C.}~\bibnamefont {Jin}}, \bibinfo
  {author} {\bibfnamefont {S.}~\bibnamefont {Guo}}, \bibinfo {author}
  {\bibfnamefont {F.}~\bibnamefont {Ning}}, \bibinfo {author} {\bibfnamefont
  {R.~M.}\ \bibnamefont {Fernandes}}, \bibinfo {author} {\bibfnamefont
  {R.}~\bibnamefont {De~Renzi}}, \bibinfo {author} {\bibfnamefont
  {P.}~\bibnamefont {Dai}}, \ and\ \bibinfo {author} {\bibfnamefont {Y.~J.}\
  \bibnamefont {Uemura}},\ }\bibfield  {title} {\bibinfo {title}
  {{Disentangling superconducting and magnetic orders in
  ${\mathrm{NaFe}}_{1\ensuremath{-}x}{\mathrm{Ni}}_{x}\mathrm{As}$ using muon
  spin rotation}},\ }\href {\doibase 10.1103/PhysRevB.97.224508} {\bibfield
  {journal} {\bibinfo  {journal} {Phys. Rev. B}\ }\textbf {\bibinfo {volume}
  {97}},\ \bibinfo {pages} {224508} (\bibinfo {year} {2018})}\BibitemShut
  {NoStop}%
\bibitem [{\citenamefont {Hashimoto}\ \emph {et~al.}(2012)\citenamefont
  {Hashimoto}, \citenamefont {Cho}, \citenamefont {Shibauchi}, \citenamefont
  {Kasahara}, \citenamefont {Mizukami}, \citenamefont {Katsumata},
  \citenamefont {Tsuruhara}, \citenamefont {Terashima}, \citenamefont {Ikeda},
  \citenamefont {Tanatar}, \citenamefont {Kitano}, \citenamefont {Salovich},
  \citenamefont {Giannetta}, \citenamefont {Walmsley}, \citenamefont
  {Carrington}, \citenamefont {Prozorov},\ and\ \citenamefont
  {Matsuda}}]{Hashimoto-S(2012)}%
  \BibitemOpen
  \bibfield  {author} {\bibinfo {author} {\bibfnamefont {K.}~\bibnamefont
  {Hashimoto}}, \bibinfo {author} {\bibfnamefont {K.}~\bibnamefont {Cho}},
  \bibinfo {author} {\bibfnamefont {T.}~\bibnamefont {Shibauchi}}, \bibinfo
  {author} {\bibfnamefont {S.}~\bibnamefont {Kasahara}}, \bibinfo {author}
  {\bibfnamefont {Y.}~\bibnamefont {Mizukami}}, \bibinfo {author}
  {\bibfnamefont {R.}~\bibnamefont {Katsumata}}, \bibinfo {author}
  {\bibfnamefont {Y.}~\bibnamefont {Tsuruhara}}, \bibinfo {author}
  {\bibfnamefont {T.}~\bibnamefont {Terashima}}, \bibinfo {author}
  {\bibfnamefont {H.}~\bibnamefont {Ikeda}}, \bibinfo {author} {\bibfnamefont
  {M.~A.}\ \bibnamefont {Tanatar}}, \bibinfo {author} {\bibfnamefont
  {H.}~\bibnamefont {Kitano}}, \bibinfo {author} {\bibfnamefont
  {N.}~\bibnamefont {Salovich}}, \bibinfo {author} {\bibfnamefont {R.~W.}\
  \bibnamefont {Giannetta}}, \bibinfo {author} {\bibfnamefont {P.}~\bibnamefont
  {Walmsley}}, \bibinfo {author} {\bibfnamefont {A.}~\bibnamefont
  {Carrington}}, \bibinfo {author} {\bibfnamefont {R.}~\bibnamefont
  {Prozorov}}, \ and\ \bibinfo {author} {\bibfnamefont {Y.}~\bibnamefont
  {Matsuda}},\ }\bibfield  {title} {\bibinfo {title} {{A Sharp Peak of the
  Zero-Temperature Penetration Depth at Optimal Composition in
  BaFe$_2$(As$_{1-x}$P$_x$)$_2$}},\ }\href {\doibase 10.1126/science.1219821}
  {\bibfield  {journal} {\bibinfo  {journal} {Science}\ }\textbf {\bibinfo
  {volume} {336}},\ \bibinfo {pages} {1554} (\bibinfo {year}
  {2012})}\BibitemShut {NoStop}%
\bibitem [{\citenamefont {Shibauchi}\ \emph {et~al.}(2014)\citenamefont
  {Shibauchi}, \citenamefont {Carrington},\ and\ \citenamefont
  {Matsuda}}]{Matsuda-ARCMP(2014)}%
  \BibitemOpen
  \bibfield  {author} {\bibinfo {author} {\bibfnamefont {T.}~\bibnamefont
  {Shibauchi}}, \bibinfo {author} {\bibfnamefont {A.}~\bibnamefont
  {Carrington}}, \ and\ \bibinfo {author} {\bibfnamefont {Y.}~\bibnamefont
  {Matsuda}},\ }\bibfield  {title} {\bibinfo {title} {{A Quantum Critical Point
  Lying Beneath the Superconducting Dome in Iron Pnictides}},\ }\href {\doibase
  10.1146/annurev-conmatphys-031113-133921} {\bibfield  {journal} {\bibinfo
  {journal} {Annu. Rev. Condens. Matter Phys.}\ }\textbf {\bibinfo {volume}
  {5}},\ \bibinfo {pages} {113} (\bibinfo {year} {2014})}\BibitemShut {NoStop}%
\bibitem [{\citenamefont {Joshi}\ \emph {et~al.}(2020)\citenamefont {Joshi},
  \citenamefont {Nusran}, \citenamefont {Tanatar}, \citenamefont {Cho},
  \citenamefont {Bud'ko}, \citenamefont {Canfield}, \citenamefont {Fernandes},
  \citenamefont {Levchenko},\ and\ \citenamefont
  {Prozorov}}]{Prozorov-NJP(2020)}%
  \BibitemOpen
  \bibfield  {author} {\bibinfo {author} {\bibfnamefont {K.~R.}\ \bibnamefont
  {Joshi}}, \bibinfo {author} {\bibfnamefont {N.~M.}\ \bibnamefont {Nusran}},
  \bibinfo {author} {\bibfnamefont {M.~A.}\ \bibnamefont {Tanatar}}, \bibinfo
  {author} {\bibfnamefont {K.}~\bibnamefont {Cho}}, \bibinfo {author}
  {\bibfnamefont {S.~L.}\ \bibnamefont {Bud'ko}}, \bibinfo {author}
  {\bibfnamefont {P.~C.}\ \bibnamefont {Canfield}}, \bibinfo {author}
  {\bibfnamefont {R.~M.}\ \bibnamefont {Fernandes}}, \bibinfo {author}
  {\bibfnamefont {A.}~\bibnamefont {Levchenko}}, \ and\ \bibinfo {author}
  {\bibfnamefont {R.}~\bibnamefont {Prozorov}},\ }\bibfield  {title} {\bibinfo
  {title} {{Quantum phase transition inside the superconducting dome of
  Ba(Fe$_{1-x}$Co$_x$)$_2$As$_2$ from diamond-based optical magnetometry}},\
  }\href {\doibase 10.1088/1367-2630/ab85a9} {\bibfield  {journal} {\bibinfo
  {journal} {New J. Phys.}\ }\textbf {\bibinfo {volume} {22}},\ \bibinfo
  {pages} {053037} (\bibinfo {year} {2020})}\BibitemShut {NoStop}%
\bibitem [{\citenamefont {Nicklas}\ \emph {et~al.}(2007)\citenamefont
  {Nicklas}, \citenamefont {Stockert}, \citenamefont {Park}, \citenamefont
  {Habicht}, \citenamefont {Kiefer}, \citenamefont {Pham}, \citenamefont
  {Thompson}, \citenamefont {Fisk},\ and\ \citenamefont
  {Steglich}}]{Nicklas2007}%
  \BibitemOpen
  \bibfield  {author} {\bibinfo {author} {\bibfnamefont {M.}~\bibnamefont
  {Nicklas}}, \bibinfo {author} {\bibfnamefont {O.}~\bibnamefont {Stockert}},
  \bibinfo {author} {\bibfnamefont {T.}~\bibnamefont {Park}}, \bibinfo {author}
  {\bibfnamefont {K.}~\bibnamefont {Habicht}}, \bibinfo {author} {\bibfnamefont
  {K.}~\bibnamefont {Kiefer}}, \bibinfo {author} {\bibfnamefont {L.~D.}\
  \bibnamefont {Pham}}, \bibinfo {author} {\bibfnamefont {J.~D.}\ \bibnamefont
  {Thompson}}, \bibinfo {author} {\bibfnamefont {Z.}~\bibnamefont {Fisk}}, \
  and\ \bibinfo {author} {\bibfnamefont {F.}~\bibnamefont {Steglich}},\
  }\bibfield  {title} {\bibinfo {title} {{Magnetic structure of Cd-doped
  $\mathrm{Ce}\mathrm{Co}{\mathrm{In}}_{5}$}},\ }\href {\doibase
  10.1103/PhysRevB.76.052401} {\bibfield  {journal} {\bibinfo  {journal} {Phys.
  Rev. B}\ }\textbf {\bibinfo {volume} {76}},\ \bibinfo {pages} {052401}
  (\bibinfo {year} {2007})}\BibitemShut {NoStop}%
\bibitem [{\citenamefont {Rosa}\ \emph {et~al.}(2017)\citenamefont {Rosa},
  \citenamefont {Kang}, \citenamefont {Luo}, \citenamefont {Wakeham},
  \citenamefont {Bauer}, \citenamefont {Ronning}, \citenamefont {Fisk},
  \citenamefont {Fernandes},\ and\ \citenamefont {Thompson}}]{Rosa2017}%
  \BibitemOpen
  \bibfield  {author} {\bibinfo {author} {\bibfnamefont {P.~F.~S.}\
  \bibnamefont {Rosa}}, \bibinfo {author} {\bibfnamefont {J.}~\bibnamefont
  {Kang}}, \bibinfo {author} {\bibfnamefont {Y.}~\bibnamefont {Luo}}, \bibinfo
  {author} {\bibfnamefont {N.}~\bibnamefont {Wakeham}}, \bibinfo {author}
  {\bibfnamefont {E.~D.}\ \bibnamefont {Bauer}}, \bibinfo {author}
  {\bibfnamefont {F.}~\bibnamefont {Ronning}}, \bibinfo {author} {\bibfnamefont
  {Z.}~\bibnamefont {Fisk}}, \bibinfo {author} {\bibfnamefont {R.~M.}\
  \bibnamefont {Fernandes}}, \ and\ \bibinfo {author} {\bibfnamefont {J.~D.}\
  \bibnamefont {Thompson}},\ }\bibfield  {title} {\bibinfo {title} {{Competing
  magnetic orders in the superconducting state of heavy-fermion
  $\mathrm{Ce}\mathrm{Rh}{\mathrm{In}}_{5}$}},\ }\href {\doibase
  10.1073/pnas.1703016114} {\bibfield  {journal} {\bibinfo  {journal} {Proc.
  Natl. Acad. Sci. (U.S.A.)}\ }\textbf {\bibinfo {volume} {114}},\ \bibinfo
  {pages} {5384} (\bibinfo {year} {2017})}\BibitemShut {NoStop}%
\bibitem [{\citenamefont {Levchenko}\ \emph {et~al.}(2013)\citenamefont
  {Levchenko}, \citenamefont {Vavilov}, \citenamefont {Khodas},\ and\
  \citenamefont {Chubukov}}]{Chubukov-PRL(2013)}%
  \BibitemOpen
  \bibfield  {author} {\bibinfo {author} {\bibfnamefont {A.}~\bibnamefont
  {Levchenko}}, \bibinfo {author} {\bibfnamefont {M.~G.}\ \bibnamefont
  {Vavilov}}, \bibinfo {author} {\bibfnamefont {M.}~\bibnamefont {Khodas}}, \
  and\ \bibinfo {author} {\bibfnamefont {A.~V.}\ \bibnamefont {Chubukov}},\
  }\bibfield  {title} {\bibinfo {title} {{Enhancement of the London Penetration
  Depth in Pnictides at the Onset of Spin-Density-Wave Order under
  Superconducting Dome}},\ }\href {\doibase 10.1103/PhysRevLett.110.177003}
  {\bibfield  {journal} {\bibinfo  {journal} {Phys. Rev. Lett.}\ }\textbf
  {\bibinfo {volume} {110}},\ \bibinfo {pages} {177003} (\bibinfo {year}
  {2013})}\BibitemShut {NoStop}%
\bibitem [{\citenamefont {Chowdhury}\ \emph {et~al.}(2013)\citenamefont
  {Chowdhury}, \citenamefont {Swingle}, \citenamefont {Berg},\ and\
  \citenamefont {Sachdev}}]{Chowdhury-PRL(2013)}%
  \BibitemOpen
  \bibfield  {author} {\bibinfo {author} {\bibfnamefont {D.}~\bibnamefont
  {Chowdhury}}, \bibinfo {author} {\bibfnamefont {B.}~\bibnamefont {Swingle}},
  \bibinfo {author} {\bibfnamefont {E.}~\bibnamefont {Berg}}, \ and\ \bibinfo
  {author} {\bibfnamefont {S.}~\bibnamefont {Sachdev}},\ }\bibfield  {title}
  {\bibinfo {title} {{Singularity of the London Penetration Depth at Quantum
  Critical Points in Superconductors}},\ }\href {\doibase
  10.1103/PhysRevLett.111.157004} {\bibfield  {journal} {\bibinfo  {journal}
  {Phys. Rev. Lett.}\ }\textbf {\bibinfo {volume} {111}},\ \bibinfo {pages}
  {157004} (\bibinfo {year} {2013})}\BibitemShut {NoStop}%
\bibitem [{\citenamefont {Nomoto}\ and\ \citenamefont
  {Ikeda}(2013)}]{Ikeda2013}%
  \BibitemOpen
  \bibfield  {author} {\bibinfo {author} {\bibfnamefont {T.}~\bibnamefont
  {Nomoto}}\ and\ \bibinfo {author} {\bibfnamefont {H.}~\bibnamefont {Ikeda}},\
  }\bibfield  {title} {\bibinfo {title} {{Effect of Magnetic Criticality and
  Fermi-Surface Topology on the Magnetic Penetration Depth}},\ }\href {\doibase
  10.1103/PhysRevLett.111.167001} {\bibfield  {journal} {\bibinfo  {journal}
  {Phys. Rev. Lett.}\ }\textbf {\bibinfo {volume} {111}},\ \bibinfo {pages}
  {167001} (\bibinfo {year} {2013})}\BibitemShut {NoStop}%
\bibitem [{\citenamefont {Chowdhury}\ \emph {et~al.}(2015)\citenamefont
  {Chowdhury}, \citenamefont {Orenstein}, \citenamefont {Sachdev},\ and\
  \citenamefont {Senthil}}]{Chowdhury2015}%
  \BibitemOpen
  \bibfield  {author} {\bibinfo {author} {\bibfnamefont {D.}~\bibnamefont
  {Chowdhury}}, \bibinfo {author} {\bibfnamefont {J.}~\bibnamefont
  {Orenstein}}, \bibinfo {author} {\bibfnamefont {S.}~\bibnamefont {Sachdev}},
  \ and\ \bibinfo {author} {\bibfnamefont {T.}~\bibnamefont {Senthil}},\
  }\bibfield  {title} {\bibinfo {title} {{Phase transition beneath the
  superconducting dome in
  ${\text{BaFe}}_{2}{({\text{As}}_{1\ensuremath{-}x}{\text{P}}_{x})}_{2}$}},\
  }\href {\doibase 10.1103/PhysRevB.92.081113} {\bibfield  {journal} {\bibinfo
  {journal} {Phys. Rev. B}\ }\textbf {\bibinfo {volume} {92}},\ \bibinfo
  {pages} {081113} (\bibinfo {year} {2015})}\BibitemShut {NoStop}%
\bibitem [{\citenamefont {Dzero}\ \emph {et~al.}(2015)\citenamefont {Dzero},
  \citenamefont {Khodas}, \citenamefont {Klironomos}, \citenamefont {Vavilov},\
  and\ \citenamefont {Levchenko}}]{Dzero2015}%
  \BibitemOpen
  \bibfield  {author} {\bibinfo {author} {\bibfnamefont {M.}~\bibnamefont
  {Dzero}}, \bibinfo {author} {\bibfnamefont {M.}~\bibnamefont {Khodas}},
  \bibinfo {author} {\bibfnamefont {A.~D.}\ \bibnamefont {Klironomos}},
  \bibinfo {author} {\bibfnamefont {M.~G.}\ \bibnamefont {Vavilov}}, \ and\
  \bibinfo {author} {\bibfnamefont {A.}~\bibnamefont {Levchenko}},\ }\bibfield
  {title} {\bibinfo {title} {{Magnetic penetration depth in disordered
  iron-based superconductors}},\ }\href {\doibase 10.1103/PhysRevB.92.144501}
  {\bibfield  {journal} {\bibinfo  {journal} {Phys. Rev. B}\ }\textbf {\bibinfo
  {volume} {92}},\ \bibinfo {pages} {144501} (\bibinfo {year}
  {2015})}\BibitemShut {NoStop}%
\bibitem [{\citenamefont {Hirschfeld}\ \emph {et~al.}(2011)\citenamefont
  {Hirschfeld}, \citenamefont {Korshunov},\ and\ \citenamefont
  {Mazin}}]{Hirschfeld2011}%
  \BibitemOpen
  \bibfield  {author} {\bibinfo {author} {\bibfnamefont {P.~J.}\ \bibnamefont
  {Hirschfeld}}, \bibinfo {author} {\bibfnamefont {M.~M.}\ \bibnamefont
  {Korshunov}}, \ and\ \bibinfo {author} {\bibfnamefont {I.~I.}\ \bibnamefont
  {Mazin}},\ }\bibfield  {title} {\bibinfo {title} {{Gap symmetry and structure
  of Fe-based superconductors}},\ }\href {\doibase
  10.1088/0034-4885/74/12/124508} {\bibfield  {journal} {\bibinfo  {journal}
  {Rep. Prog. Phys.}\ }\textbf {\bibinfo {volume} {74}},\ \bibinfo {pages}
  {124508} (\bibinfo {year} {2011})}\BibitemShut {NoStop}%
\bibitem [{\citenamefont {Chubukov}(2012)}]{Chubukov-ARCMP(2012)}%
  \BibitemOpen
  \bibfield  {author} {\bibinfo {author} {\bibfnamefont {A.}~\bibnamefont
  {Chubukov}},\ }\bibfield  {title} {\bibinfo {title} {{Pairing Mechanism in
  Fe-Based Superconductors}},\ }\href {\doibase
  10.1146/annurev-conmatphys-020911-125055} {\bibfield  {journal} {\bibinfo
  {journal} {Annu. Rev. Condens. Matter Phys.}\ }\textbf {\bibinfo {volume}
  {3}},\ \bibinfo {pages} {57} (\bibinfo {year} {2012})}\BibitemShut {NoStop}%
\bibitem [{\citenamefont {Li}\ \emph {et~al.}(2018)\citenamefont {Li},
  \citenamefont {Liu}, \citenamefont {Fu}, \citenamefont {Chen}, \citenamefont
  {Yang},\ and\ \citenamefont {Yang}}]{Yifeng2018}%
  \BibitemOpen
  \bibfield  {author} {\bibinfo {author} {\bibfnamefont {Y.}~\bibnamefont
  {Li}}, \bibinfo {author} {\bibfnamefont {M.}~\bibnamefont {Liu}}, \bibinfo
  {author} {\bibfnamefont {Z.}~\bibnamefont {Fu}}, \bibinfo {author}
  {\bibfnamefont {X.}~\bibnamefont {Chen}}, \bibinfo {author} {\bibfnamefont
  {F.}~\bibnamefont {Yang}}, \ and\ \bibinfo {author} {\bibfnamefont {Y.-f.}\
  \bibnamefont {Yang}},\ }\bibfield  {title} {\bibinfo {title} {{Gap Symmetry
  of the Heavy Fermion Superconductor ${\mathrm{CeCu}}_{2}{\mathrm{Si}}_{2}$ at
  Ambient Pressure}},\ }\href {\doibase 10.1103/PhysRevLett.120.217001}
  {\bibfield  {journal} {\bibinfo  {journal} {Phys. Rev. Lett.}\ }\textbf
  {\bibinfo {volume} {120}},\ \bibinfo {pages} {217001} (\bibinfo {year}
  {2018})}\BibitemShut {NoStop}%
\bibitem [{\citenamefont {Kos}\ \emph {et~al.}(2003)\citenamefont {Kos},
  \citenamefont {Martin},\ and\ \citenamefont {Varma}}]{Varma2003}%
  \BibitemOpen
  \bibfield  {author} {\bibinfo {author} {\bibfnamefont {{\ifmmode
  \check{S}\else \v{S}}.}~\bibnamefont {Kos}}, \bibinfo {author} {\bibfnamefont
  {I.}~\bibnamefont {Martin}}, \ and\ \bibinfo {author} {\bibfnamefont {C.~M.}\
  \bibnamefont {Varma}},\ }\bibfield  {title} {\bibinfo {title} {{Specific heat
  at the transition in a superconductor with fluctuating magnetic moments}},\
  }\href {\doibase 10.1103/PhysRevB.68.052507} {\bibfield  {journal} {\bibinfo
  {journal} {Phys. Rev. B}\ }\textbf {\bibinfo {volume} {68}},\ \bibinfo
  {pages} {052507} (\bibinfo {year} {2003})}\BibitemShut {NoStop}%
\bibitem [{\citenamefont {Kuzmanovski}\ \emph {et~al.}(2014)\citenamefont
  {Kuzmanovski}, \citenamefont {Levchenko}, \citenamefont {Khodas},\ and\
  \citenamefont {Vavilov}}]{Vavilov-PRB(2014)}%
  \BibitemOpen
  \bibfield  {author} {\bibinfo {author} {\bibfnamefont {D.}~\bibnamefont
  {Kuzmanovski}}, \bibinfo {author} {\bibfnamefont {A.}~\bibnamefont
  {Levchenko}}, \bibinfo {author} {\bibfnamefont {M.}~\bibnamefont {Khodas}}, \
  and\ \bibinfo {author} {\bibfnamefont {M.~G.}\ \bibnamefont {Vavilov}},\
  }\bibfield  {title} {\bibinfo {title} {{Effect of spin-density wave
  fluctuations on the specific heat jump in iron pnictides at the
  superconducting transition}},\ }\href {\doibase 10.1103/PhysRevB.89.144503}
  {\bibfield  {journal} {\bibinfo  {journal} {Phys. Rev. B}\ }\textbf {\bibinfo
  {volume} {89}},\ \bibinfo {pages} {144503} (\bibinfo {year}
  {2014})}\BibitemShut {NoStop}%
\bibitem [{\citenamefont {Walmsley}\ \emph {et~al.}(2013)\citenamefont
  {Walmsley}, \citenamefont {Putzke}, \citenamefont {Malone}, \citenamefont
  {Guillam\'on}, \citenamefont {Vignolles}, \citenamefont {Proust},
  \citenamefont {Badoux}, \citenamefont {Coldea}, \citenamefont {Watson},
  \citenamefont {Kasahara}, \citenamefont {Mizukami}, \citenamefont
  {Shibauchi}, \citenamefont {Matsuda},\ and\ \citenamefont
  {Carrington}}]{Carrington-PRL(2013)}%
  \BibitemOpen
  \bibfield  {author} {\bibinfo {author} {\bibfnamefont {P.}~\bibnamefont
  {Walmsley}}, \bibinfo {author} {\bibfnamefont {C.}~\bibnamefont {Putzke}},
  \bibinfo {author} {\bibfnamefont {L.}~\bibnamefont {Malone}}, \bibinfo
  {author} {\bibfnamefont {I.}~\bibnamefont {Guillam\'on}}, \bibinfo {author}
  {\bibfnamefont {D.}~\bibnamefont {Vignolles}}, \bibinfo {author}
  {\bibfnamefont {C.}~\bibnamefont {Proust}}, \bibinfo {author} {\bibfnamefont
  {S.}~\bibnamefont {Badoux}}, \bibinfo {author} {\bibfnamefont {A.~I.}\
  \bibnamefont {Coldea}}, \bibinfo {author} {\bibfnamefont {M.~D.}\
  \bibnamefont {Watson}}, \bibinfo {author} {\bibfnamefont {S.}~\bibnamefont
  {Kasahara}}, \bibinfo {author} {\bibfnamefont {Y.}~\bibnamefont {Mizukami}},
  \bibinfo {author} {\bibfnamefont {T.}~\bibnamefont {Shibauchi}}, \bibinfo
  {author} {\bibfnamefont {Y.}~\bibnamefont {Matsuda}}, \ and\ \bibinfo
  {author} {\bibfnamefont {A.}~\bibnamefont {Carrington}},\ }\bibfield  {title}
  {\bibinfo {title} {{Quasiparticle Mass Enhancement Close to the Quantum
  Critical Point in
  ${\mathrm{BaFe}}_{2}({\mathrm{As}}_{1\ensuremath{-}x}{\mathrm{P}}_{x}{)}_{2}$}},\
  }\href {\doibase 10.1103/PhysRevLett.110.257002} {\bibfield  {journal}
  {\bibinfo  {journal} {Phys. Rev. Lett.}\ }\textbf {\bibinfo {volume} {110}},\
  \bibinfo {pages} {257002} (\bibinfo {year} {2013})}\BibitemShut {NoStop}%
\bibitem [{\citenamefont {Chubukov}\ \emph {et~al.}(1994)\citenamefont
  {Chubukov}, \citenamefont {Sachdev},\ and\ \citenamefont
  {Ye}}]{Chubukov-PRB(1994)}%
  \BibitemOpen
  \bibfield  {author} {\bibinfo {author} {\bibfnamefont {A.~V.}\ \bibnamefont
  {Chubukov}}, \bibinfo {author} {\bibfnamefont {S.}~\bibnamefont {Sachdev}}, \
  and\ \bibinfo {author} {\bibfnamefont {J.}~\bibnamefont {Ye}},\ }\bibfield
  {title} {\bibinfo {title} {{Theory of two-dimensional quantum Heisenberg
  antiferromagnets with a nearly critical ground state}},\ }\href {\doibase
  10.1103/PhysRevB.49.11919} {\bibfield  {journal} {\bibinfo  {journal} {Phys.
  Rev. B}\ }\textbf {\bibinfo {volume} {49}},\ \bibinfo {pages} {11919}
  (\bibinfo {year} {1994})}\BibitemShut {NoStop}%
\bibitem [{\citenamefont {Fernandes}\ and\ \citenamefont
  {Chubukov}(2017)}]{Fernandes-RPP(2017)}%
  \BibitemOpen
  \bibfield  {author} {\bibinfo {author} {\bibfnamefont {R.~M.}\ \bibnamefont
  {Fernandes}}\ and\ \bibinfo {author} {\bibfnamefont {A.~V.}\ \bibnamefont
  {Chubukov}},\ }\bibfield  {title} {\bibinfo {title} {{Low-energy microscopic
  models for iron-based superconductors: a review}},\ }\href
  {http://stacks.iop.org/0034-4885/80/i=1/a=014503} {\bibfield  {journal}
  {\bibinfo  {journal} {Rep. Prog. Phys.}\ }\textbf {\bibinfo {volume} {80}},\
  \bibinfo {pages} {014503} (\bibinfo {year} {2017})}\BibitemShut {NoStop}%
\bibitem [{\citenamefont {Chubukov}\ \emph {et~al.}(2008)\citenamefont
  {Chubukov}, \citenamefont {Efremov},\ and\ \citenamefont
  {Eremin}}]{Chubukov-PRB(2008)}%
  \BibitemOpen
  \bibfield  {author} {\bibinfo {author} {\bibfnamefont {A.~V.}\ \bibnamefont
  {Chubukov}}, \bibinfo {author} {\bibfnamefont {D.~V.}\ \bibnamefont
  {Efremov}}, \ and\ \bibinfo {author} {\bibfnamefont {I.}~\bibnamefont
  {Eremin}},\ }\bibfield  {title} {\bibinfo {title} {{Magnetism,
  superconductivity, and pairing symmetry in iron-based superconductors}},\
  }\href {\doibase 10.1103/PhysRevB.78.134512} {\bibfield  {journal} {\bibinfo
  {journal} {Phys. Rev. B}\ }\textbf {\bibinfo {volume} {78}},\ \bibinfo
  {pages} {134512} (\bibinfo {year} {2008})}\BibitemShut {NoStop}%
\bibitem [{\citenamefont {Vavilov}\ \emph {et~al.}(2010)\citenamefont
  {Vavilov}, \citenamefont {Chubukov},\ and\ \citenamefont
  {Vorontsov}}]{Vavilov-SST(2010)}%
  \BibitemOpen
  \bibfield  {author} {\bibinfo {author} {\bibfnamefont {M.~G.}\ \bibnamefont
  {Vavilov}}, \bibinfo {author} {\bibfnamefont {A.~V.}\ \bibnamefont
  {Chubukov}}, \ and\ \bibinfo {author} {\bibfnamefont {A.~B.}\ \bibnamefont
  {Vorontsov}},\ }\bibfield  {title} {\bibinfo {title} {{Coexistence between
  superconducting and spin density wave states in iron-based superconductors:
  Ginzburg{\textendash}Landau analysis}},\ }\href {\doibase
  10.1088/0953-2048/23/5/054011} {\bibfield  {journal} {\bibinfo  {journal}
  {Supercond. Sci. and Technol.}\ }\textbf {\bibinfo {volume} {23}},\ \bibinfo
  {pages} {054011} (\bibinfo {year} {2010})}\BibitemShut {NoStop}%
\bibitem [{\citenamefont {Millis}(1993)}]{Millis-PRB(1993)}%
  \BibitemOpen
  \bibfield  {author} {\bibinfo {author} {\bibfnamefont {A.~J.}\ \bibnamefont
  {Millis}},\ }\bibfield  {title} {\bibinfo {title} {{Effect of a nonzero
  temperature on quantum critical points in itinerant fermion systems}},\
  }\href {\doibase 10.1103/PhysRevB.48.7183} {\bibfield  {journal} {\bibinfo
  {journal} {Phys. Rev. B}\ }\textbf {\bibinfo {volume} {48}},\ \bibinfo
  {pages} {7183} (\bibinfo {year} {1993})}\BibitemShut {NoStop}%
\bibitem [{\citenamefont {Abanov}\ and\ \citenamefont
  {Chubukov}(2004)}]{Chubukov-PRL(2004)}%
  \BibitemOpen
  \bibfield  {author} {\bibinfo {author} {\bibfnamefont {A.}~\bibnamefont
  {Abanov}}\ and\ \bibinfo {author} {\bibfnamefont {A.}~\bibnamefont
  {Chubukov}},\ }\bibfield  {title} {\bibinfo {title} {{Anomalous Scaling at
  the Quantum Critical Point in Itinerant Antiferromagnets}},\ }\href {\doibase
  10.1103/PhysRevLett.93.255702} {\bibfield  {journal} {\bibinfo  {journal}
  {Phys. Rev. Lett.}\ }\textbf {\bibinfo {volume} {93}},\ \bibinfo {pages}
  {255702} (\bibinfo {year} {2004})}\BibitemShut {NoStop}%
\bibitem [{\citenamefont {Chubukov}\ \emph {et~al.}(2005)\citenamefont
  {Chubukov}, \citenamefont {Maslov}, \citenamefont {Gangadharaiah},\ and\
  \citenamefont {Glazman}}]{Chubukov-PRB(2005)}%
  \BibitemOpen
  \bibfield  {author} {\bibinfo {author} {\bibfnamefont {A.~V.}\ \bibnamefont
  {Chubukov}}, \bibinfo {author} {\bibfnamefont {D.~L.}\ \bibnamefont
  {Maslov}}, \bibinfo {author} {\bibfnamefont {S.}~\bibnamefont
  {Gangadharaiah}}, \ and\ \bibinfo {author} {\bibfnamefont {L.~I.}\
  \bibnamefont {Glazman}},\ }\bibfield  {title} {\bibinfo {title} {{Singular
  perturbation theory for interacting fermions in two dimensions}},\ }\href
  {\doibase 10.1103/PhysRevB.71.205112} {\bibfield  {journal} {\bibinfo
  {journal} {Phys. Rev. B}\ }\textbf {\bibinfo {volume} {71}},\ \bibinfo
  {pages} {205112} (\bibinfo {year} {2005})}\BibitemShut {NoStop}%
\bibitem [{\citenamefont {Luttinger}\ and\ \citenamefont
  {Ward}(1960)}]{LW_1960}%
  \BibitemOpen
  \bibfield  {author} {\bibinfo {author} {\bibfnamefont {J.~M.}\ \bibnamefont
  {Luttinger}}\ and\ \bibinfo {author} {\bibfnamefont {J.~C.}\ \bibnamefont
  {Ward}},\ }\bibfield  {title} {\bibinfo {title} {{Ground-State Energy of a
  Many-Fermion System. II}},\ }\href {\doibase 10.1103/PhysRev.118.1417}
  {\bibfield  {journal} {\bibinfo  {journal} {Phys. Rev.}\ }\textbf {\bibinfo
  {volume} {118}},\ \bibinfo {pages} {1417} (\bibinfo {year}
  {1960})}\BibitemShut {NoStop}%
\bibitem [{\citenamefont {Eliashberg}(1960)}]{Eliashberg_1960}%
  \BibitemOpen
  \bibfield  {author} {\bibinfo {author} {\bibfnamefont {G.~M.}\ \bibnamefont
  {Eliashberg}},\ }\bibfield  {title} {\bibinfo {title} {{Interactions between
  Electrons and Lattice Vibrations in a Superconductor}},\ }\href
  {http://www.jetp.ac.ru/cgi-bin/e/index/e/11/3/p696?a=list} {\bibfield
  {journal} {\bibinfo  {journal} {Sov. Phys. JETP}\ }\textbf {\bibinfo {volume}
  {11}},\ \bibinfo {pages} {696} (\bibinfo {year} {1960})}\BibitemShut
  {NoStop}%
\bibitem [{\citenamefont {Haslinger}\ and\ \citenamefont
  {Chubukov}(2003)}]{Haslinger2003}%
  \BibitemOpen
  \bibfield  {author} {\bibinfo {author} {\bibfnamefont {R.}~\bibnamefont
  {Haslinger}}\ and\ \bibinfo {author} {\bibfnamefont {A.~V.}\ \bibnamefont
  {Chubukov}},\ }\bibfield  {title} {\bibinfo {title} {{Condensation energy in
  strongly coupled superconductors}},\ }\href {\doibase
  10.1103/PhysRevB.68.214508} {\bibfield  {journal} {\bibinfo  {journal} {Phys.
  Rev. B}\ }\textbf {\bibinfo {volume} {68}},\ \bibinfo {pages} {214508}
  (\bibinfo {year} {2003})}\BibitemShut {NoStop}%
\bibitem [{\citenamefont {Bardeen}\ \emph {et~al.}(1957)\citenamefont
  {Bardeen}, \citenamefont {Cooper},\ and\ \citenamefont
  {Schrieffer}}]{Bardeen-PR(1957)}%
  \BibitemOpen
  \bibfield  {author} {\bibinfo {author} {\bibfnamefont {J.}~\bibnamefont
  {Bardeen}}, \bibinfo {author} {\bibfnamefont {L.~N.}\ \bibnamefont {Cooper}},
  \ and\ \bibinfo {author} {\bibfnamefont {J.~R.}\ \bibnamefont {Schrieffer}},\
  }\bibfield  {title} {\bibinfo {title} {{Theory of Superconductivity}},\
  }\href {\doibase 10.1103/PhysRev.108.1175} {\bibfield  {journal} {\bibinfo
  {journal} {Phys. Rev.}\ }\textbf {\bibinfo {volume} {108}},\ \bibinfo {pages}
  {1175} (\bibinfo {year} {1957})}\BibitemShut {NoStop}%
\bibitem [{\citenamefont {Sachdev}(1993)}]{Sachdev-PLB(1993)}%
  \BibitemOpen
  \bibfield  {author} {\bibinfo {author} {\bibfnamefont {S.}~\bibnamefont
  {Sachdev}},\ }\bibfield  {title} {\bibinfo {title} {{Polylogarithm identities
  in a conformal field theory in three dimensions}},\ }\href {\doibase
  https://doi.org/10.1016/0370-2693(93)90935-B} {\bibfield  {journal} {\bibinfo
   {journal} {Phys. Lett. B}\ }\textbf {\bibinfo {volume} {309}},\ \bibinfo
  {pages} {285} (\bibinfo {year} {1993})}\BibitemShut {NoStop}%
\bibitem [{\citenamefont {Ding}\ \emph {et~al.}(2018)\citenamefont {Ding},
  \citenamefont {Meier}, \citenamefont {Cui}, \citenamefont {Xu}, \citenamefont
  {B\"ohmer}, \citenamefont {Bud'ko}, \citenamefont {Canfield},\ and\
  \citenamefont {Furukawa}}]{Furukawa-PRL(2018)}%
  \BibitemOpen
  \bibfield  {author} {\bibinfo {author} {\bibfnamefont {Q.-P.}\ \bibnamefont
  {Ding}}, \bibinfo {author} {\bibfnamefont {W.~R.}\ \bibnamefont {Meier}},
  \bibinfo {author} {\bibfnamefont {J.}~\bibnamefont {Cui}}, \bibinfo {author}
  {\bibfnamefont {M.}~\bibnamefont {Xu}}, \bibinfo {author} {\bibfnamefont
  {A.~E.}\ \bibnamefont {B\"ohmer}}, \bibinfo {author} {\bibfnamefont {S.~L.}\
  \bibnamefont {Bud'ko}}, \bibinfo {author} {\bibfnamefont {P.~C.}\
  \bibnamefont {Canfield}}, \ and\ \bibinfo {author} {\bibfnamefont
  {Y.}~\bibnamefont {Furukawa}},\ }\bibfield  {title} {\bibinfo {title}
  {{Hedgehog Spin-Vortex Crystal Antiferromagnetic Quantum Criticality in
  $\mathrm{CaK}({\mathrm{Fe}}_{1\ensuremath{-}x}{\mathrm{Ni}}_{x}{)}_{4}{\mathrm{As}}_{4}$
  Revealed by NMR}},\ }\href {\doibase 10.1103/PhysRevLett.121.137204}
  {\bibfield  {journal} {\bibinfo  {journal} {Phys. Rev. Lett.}\ }\textbf
  {\bibinfo {volume} {121}},\ \bibinfo {pages} {137204} (\bibinfo {year}
  {2018})}\BibitemShut {NoStop}%
\bibitem [{\citenamefont {Meier}\ \emph {et~al.}(2018)\citenamefont {Meier},
  \citenamefont {Ding}, \citenamefont {Kreyssig}, \citenamefont {Bud'ko},
  \citenamefont {Sapkota}, \citenamefont {Kothapalli}, \citenamefont {Borisov},
  \citenamefont {Valent\'{\i}}, \citenamefont {Batista}, \citenamefont {Orth},
  \citenamefont {Fernandes}, \citenamefont {Goldman}, \citenamefont {Furukawa},
  \citenamefont {B\"ohmer},\ and\ \citenamefont
  {Canfield}}]{Canfield-NPJ(2018)}%
  \BibitemOpen
  \bibfield  {author} {\bibinfo {author} {\bibfnamefont {W.~R.}\ \bibnamefont
  {Meier}}, \bibinfo {author} {\bibfnamefont {Q.-P.}\ \bibnamefont {Ding}},
  \bibinfo {author} {\bibfnamefont {A.}~\bibnamefont {Kreyssig}}, \bibinfo
  {author} {\bibfnamefont {S.~L.}\ \bibnamefont {Bud'ko}}, \bibinfo {author}
  {\bibfnamefont {A.}~\bibnamefont {Sapkota}}, \bibinfo {author} {\bibfnamefont
  {K.}~\bibnamefont {Kothapalli}}, \bibinfo {author} {\bibfnamefont
  {V.}~\bibnamefont {Borisov}}, \bibinfo {author} {\bibfnamefont
  {R.}~\bibnamefont {Valent\'{\i}}}, \bibinfo {author} {\bibfnamefont {C.~D.}\
  \bibnamefont {Batista}}, \bibinfo {author} {\bibfnamefont {P.~P.}\
  \bibnamefont {Orth}}, \bibinfo {author} {\bibfnamefont {R.~M.}\ \bibnamefont
  {Fernandes}}, \bibinfo {author} {\bibfnamefont {A.~I.}\ \bibnamefont
  {Goldman}}, \bibinfo {author} {\bibfnamefont {Y.}~\bibnamefont {Furukawa}},
  \bibinfo {author} {\bibfnamefont {A.~E.}\ \bibnamefont {B\"ohmer}}, \ and\
  \bibinfo {author} {\bibfnamefont {P.~C.}\ \bibnamefont {Canfield}},\
  }\bibfield  {title} {\bibinfo {title} {{Hedgehog spin-vortex crystal
  stabilized in a hole-doped iron-based superconductor}},\ }\href {\doibase
  10.1038/s41535-017-0076-x} {\bibfield  {journal} {\bibinfo  {journal} {npj
  Quantum Mater.}\ }\textbf {\bibinfo {volume} {3}},\ \bibinfo {pages} {5}
  (\bibinfo {year} {2018})}\BibitemShut {NoStop}%
\end{thebibliography}

%

\end{document}